\documentclass{article}
\usepackage{arxiv}

\usepackage[utf8]{inputenc}
\usepackage[english]{babel}
\usepackage{amsmath,float}
\usepackage{graphicx, comment}
\usepackage[colorinlistoftodos]{todonotes}
\usepackage{caption}
\usepackage{subcaption}
\usepackage{geometry}
\usepackage{hyperref} 
\usepackage{setspace} 
\usepackage{enumitem}
\usepackage{array,booktabs}
\usepackage{multirow}
\usepackage{amssymb}

\newcommand{\degs}{$^\circ$}

\newcommand{\abc}{$\{\alpha, \beta, \gamma\} $}

\title{Multilayered recoverable sandwich composite structures with architected core}

\author{
 V. Damodaran, A.G. Hahm and  P. Prabhakar$^*$ \\
  Dept. of Civil \& Environmental Engineering \\
  University of Wisconsin-Madison \\
  Madison, WI 53706  \vspace{0.05in} \\
  \texttt{$^*$pavana.prabhakar@wisc.edu}
}

\begin{document}

\maketitle

\begin{abstract}
    In this paper, we propose a novel design and fabrication strategy to produce architected core structures for use as the core in composite sandwich structures. A traditional foam core or honeycomb structure is lightweight and stiff, but susceptible to permanent deformation when subjected to excessive loading. Here we propose the use of an architected structure composed of arrays of hollow truncated cone unit cells that dissipate energy and exhibit structural recovery. These structures printed with a viscoelastic material rely on buckling of their sidewalls to dissipate energy and snap-back to prevent permanent deformation. We explore the mechanical response of these conical unit cells in terms of their buckling strength and post-buckling stability condition, and develop design maps for the same, by relating them to non-dimensional geometric parameters \abc{}, where $\alpha$ represents the slenderness of the curved sidewalls, $\beta$ is the angle of the sidewall to the base, and $\gamma$ represents the curvature of the sidewall. A validated finite element model is developed and used to investigate the effect of these parameters. We see that the peak buckling load is directly proportional to both $\alpha$ \& $\beta$, and is not dependent on $\gamma$ when the load is normalized by the volume of material in the curved sidewall. Interestingly, the post-buckling stability is influenced by $\gamma$, or the initial curvature of the sidewall, where a larger radius of curvature makes the structure less susceptible to exhibit structural bistability. The structures presented here are printed using a viscoelastic material, that causes them to exhibit pseudo-bistability, or a time-delayed recovery. This allows the structures to buckle and dissipate energy, and then recover to their original configurations without the need for external stimuli or energy.
\end{abstract}



\section{Introduction}
\begin{figure}[h]
     \centering
     \begin{subfigure}[b]{0.43\textwidth}
         \centering
         \includegraphics[width=0.9\textwidth]{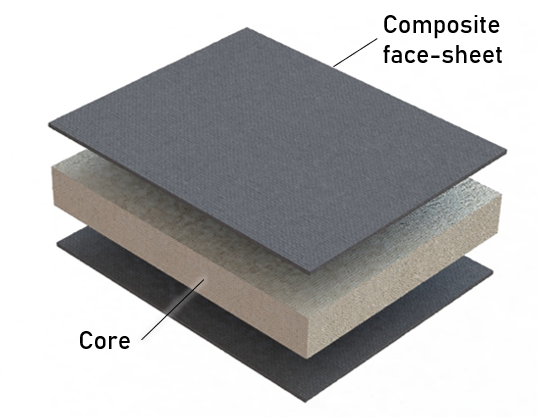}
         \caption{}
         \label{fig:traditional_foam_coare}
     \end{subfigure}
     \hfill
     \begin{subfigure}[b]{0.55\textwidth}
         \centering
         \includegraphics[width=\textwidth]{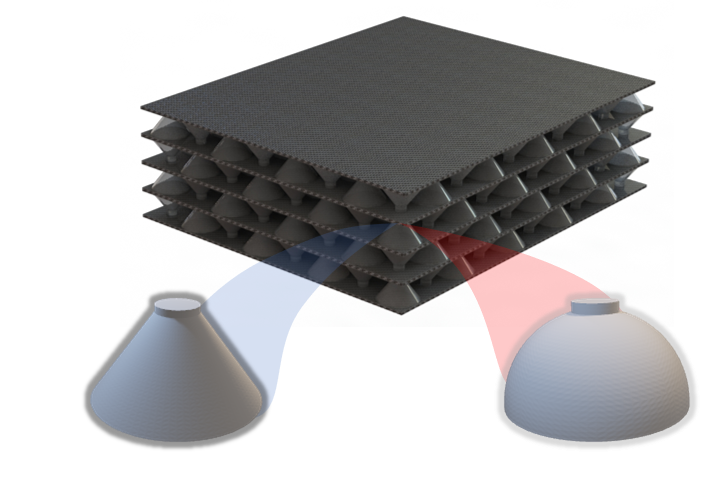}
         \caption{}
         \label{fig:architected_core}
     \end{subfigure}
     \hfill
     \begin{subfigure}[b]{\textwidth}
         \centering
         \includegraphics[width=\textwidth]{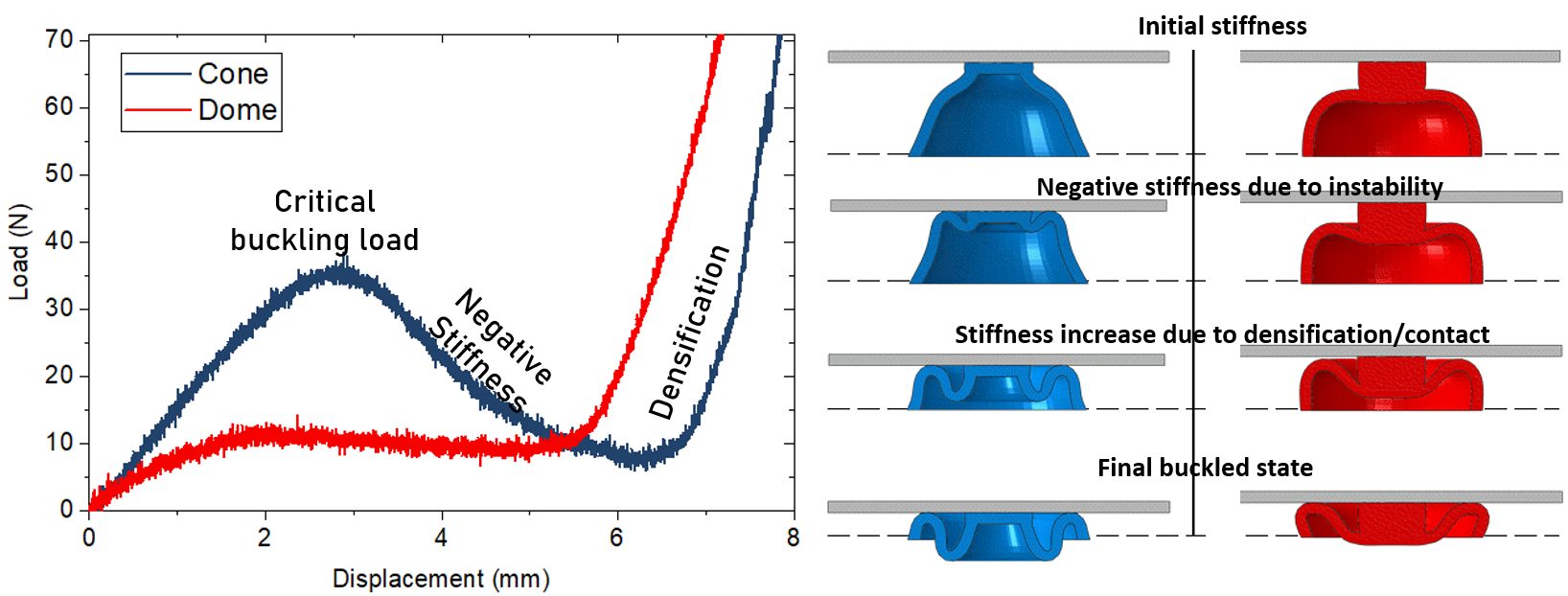}
         \caption{}
         \label{fig:initial_cone_vs_domes}
     \end{subfigure}
     \caption{a) Traditional foam core composite structures. b) Architected composite sandwich structures using viscoelastic buckling structures to dissipate energy and recover post compression. c) Preliminary studies comparing domes to cones show lower initial stiffness and peak load capacity of domes compared to their cone counterparts. Various stages of compression are highlighted in the load-displacement graph with their deformation snapshots. In a displacement controlled loading case, a negative stiffness region manifests itself which would otherwise be a \textit{snap-through} jump in a force-controlled loading case.}
\end{figure}

Composite sandwich structures offer excellent mechanical properties such as high strength and stiffness with minimal weight penalty. Traditionally, these structures consist of a lightweight core such as foams \cite{gupta_characterization_2005, breunig_dynamic_2020, rizov_indentation_2005,bai_compression_2019} or honeycomb core \cite{pathipaka_damage_2018, steeves_collapse_2004, buitrago_modelling_2010, belouettar_experimental_2009, sun_structural_2021} that are sandwiched between stiff facesheets as shown in Figure \ref{fig:traditional_foam_coare} that provide flexural rigidity. During extreme loading events such as impact, after the initial damage to the composite facesheet, the core structure absorbs most of the impact energy through inelastic methods such as crushing causing permanent localized damage to the sandwich structure \cite{meo_response_2005, breunig_dynamic_2020}. While foam cores are largely isotropic, honeycomb structures exhibit anisotropy, with their in-plane response being more compliant than out-of-plane compression response.
Researchers have explored the in-plane compression response of honeycomb structure to leverage a large loading plateau indicative of high energy dissipation \cite{cricri_honeycomb_2013, hayes_mechanics_2004}. Similar studies for hierarchical and architected structures have shown a large plateau region. \textit{In this study we aim to design an architected structure that provides a similar large plateau region but relies on buckling to dissipate energy. In addition to energy dissipation, using viscoelastic materials to fabricate the architected materials, we show structural recovery upon releasing the compressive loading.}

Structural stability is an important design consideration for the design of engineering structures. 
Buckling can alter the structural configuration leading to sudden collapse and drastically reduce the load carrying capacity of a structure. Most engineering structures are designed to avoid buckling. However, buckling can be beneficial for applications where the structural configuration can be an advantage. For example, in electrical switches where the load does not need to be applied continuously to maintain the switch in the on/off position. Such structures are called bistable structures, and usually involve a component designed to buckle to jump from one stable state to another. Other structures that undergo buckling but return to their original configuration are termed monostable. Upon removing the load, structures are designed to return to their original configuration, for example, clickers, bimetallic thermal cut-off switches, etc. From an energy consideration, the strain energy-displacement diagrams of monostable structures are monotonically increasing. On the other hand, structural bistability is characterized by having two distinct energy wells in the strain energy-displacement diagram. If a structure is pushed from one stable configuration to an other by applying an external load, upon removal of this load, the structure now resides in this new configuration, thus exhibiting bistability. This property has been explored by several researchers for impact energy absorption \cite{Shan2015, tan_novel_2019, chen_novel_2020}, switches, valves and pumps \cite{dorfmeister_static_2018, capanu_design_2000, casals-terre_snap-action_2008, rothemund_soft_2018, schomburg_design_1998, goll_microvalves_1996, uusitalo_novel_2010}, and more recently, as a mechanism for morphing structures \cite{pirrera_bistable_2010, wang_bistable_2015, zhang_bistable_2019, chen_nonlinear_2012, schultz_concept_2008}. Bistability exists in nature in the form of snapping mechanism of the Venus fly trap \cite{forterre_how_2005}. The switching from one structural configuration to another generally involves \textit{snap-through} buckling which occurs as a sudden snapping due to an instability.

While both monostable and bistable structures can exhibit snap through buckling instability, only monostable structure recover their configuration upon removal of the external load. During this period of snap-through instability, the structure exhibits negative stiffness. Upon reaching a certain minimum load response, the structure stiffens again due to a variety of reasons including densification in foams from contact or from members that were previously under compression switching to tension after buckling like in von Mises trusses. Other researchers have explored this transition from negative to positive stiffness by creating structures that are stacked together like springs in series. Such structures are used for energy dissipation purposes \cite{alturki_multistable_2019, cui_highly_2015, tan_novel_2019, restrepo_phase_2015, tan_novel_2020} due to their asymmetric load response between loading and unloading phases. Further, this introduces the concept of multi-stability, where several bistable structures arranged in series can allow the structure to exist in any combination of bistable states. These energy dissipating multistable structures need external input to reconfigure them back to their original undeformed configuration. Researchers have shown that certain external stimuli such as change in temperature can trigger recovery in viscoelastic architected materials \cite{che_viscoelastic_2018, tao_4d_2020, liu_4d_2020}. Magnetic or electric fields \cite{medina_bistable_2016, seffen_eversion_2016} can also be used to trigger the snapping action of the structure from one stable configuration to another.

An intermediate stage between monostability and bistability is a state of pseudo-bistability introduced by Santer \cite{santer_self-actuated_2010} to describe time delayed transition from bistability to monostability. Here, the snap-through buckling instability that generally occurs when bistable structures switch from one stable configuration to another is followed by a delayed \textit{snap-back} \cite{brinkmeyer_pseudo-bistable_2012} that enables structural recovery to the original configuration after a certain period of time due to viscoelastic relaxation. A viscoelastic material continues to relax and asymptotically approach a relaxed stiffness. During equilibrium in the second stable state, the material continues to relax to a point where, under certain geometric configurations, the equilibrium is lost and the structure can snap back to its first stable/undeformed configuration.

In this paper, we present an architected structure with potential application as core in a composite sandwich structure. Leveraging the energy dissipation characteristics of multi-stable structures, the architected core also introduces favorable design characteristics such as resilience due to structural recovery. Unlike other multi-stable structures that require external stimuli to trigger the snap back action, here, material relaxation of a viscoelastic material is used to push the structure back to its original configuration. The architected core structure is composed of stacked conical frusta as shown in Figure \ref{fig:architected_core}. The geometric parameter space of the cones is explored through a validated finite element model for generating design maps that can be used to produce either mono-, pseudo-, or bi-stable structures with specific peak buckling load capacities. 

\section{Motivation}
\label{se:motivation}

The goal of this study is to develop additively manufactured architected structures that can be used as  core materials in sandwich composites. Traditional core materials and structures such as foam core or honeycomb structures are good options for designing lightweight and stiff structures. These cores are known to dissipate energy under extreme conditions such as excessive compression or impact loading, where the cores experience permanent deformation that maybe detrimental to the complete structure by reducing its residual strength. In contrast to a traditional foam or honeycomb structure, a multi-layer architected core design proposed in this paper as shown in Figure~\ref{fig:architected_core} facilitates significant deformation beyond the initial peak load, yielding a larger energy dissipation during compression and other extreme loading scenarios like impact. Replacing the core structure with architected materials means the structures must be capable of supporting the structure in compression and shear loading. Additionally, in this study, the structure must be designed to undergo substantial deformation without permanent set and recover their original geometric configuration after the loading is removed. 

The architected structures presented in this paper consist of unit cells made of thin walled truncated cones and topped with a solid stub that allows for load distribution. Preliminary studies were performed to compare the performance of unit cells made with truncated cones and domes under compressive loading. The load-displacement response typically consists of three regions: 1) an initial linear regions, 2) a softening behavior after critical buckling load, and 3) stiffening due to secondary contact between the sidewalls and facesheets. The results revealed that dome shaped unit cells under-performed in terms of initial stiffness and critical load carrying capacity as compared to their conical counterparts as shown in Figure~\ref{fig:initial_cone_vs_domes}. This is attributed to the initial curvature of the sidewalls in dome unit cells that undergo bending prematurely.

\section{Methods} 
\label{se:methods}

This section describes the parametric design of the unit cell used in the generation of the architected core structure followed by the fabrication methods used to assemble a composite sandwich structure. The material properties of the 3D printed resin are obtained experimentally for use in a finite element model developed to explore the parameter space and document the load response and stability behavior of unit cells in compression and shear loading for varying parametric combinations. This is followed by the experimental validation of the finite element models, and an evaluation of the sandwich structures in bending. The methods section is followed by results and discussion.

\subsection{Unit Cell Design} \label{sse:unit_cell}
\begin{figure}[h]
    \centering
    
    \begin{subfigure}[b]{0.33\textwidth}
        \includegraphics[width=\textwidth]{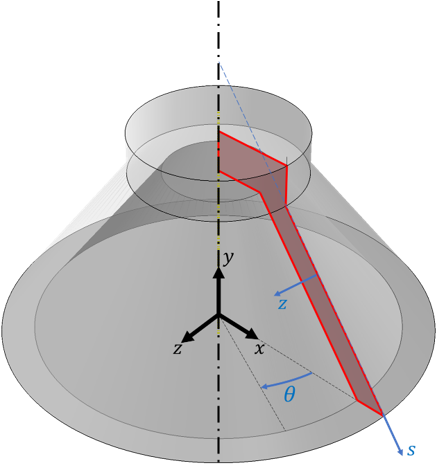}
        \centering
        \caption{}
    \end{subfigure}
    \hfill
    \begin{subfigure}[b]{0.33\textwidth}
        \centering
        \includegraphics[width=\textwidth]{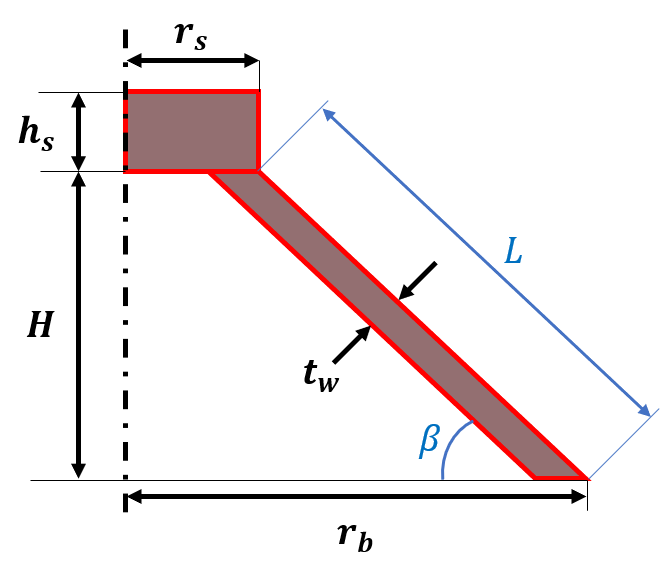}
        \caption{}
    \end{subfigure}
    \hfill
    \begin{subfigure}[b]{0.32\textwidth}
        \includegraphics[width=\textwidth]{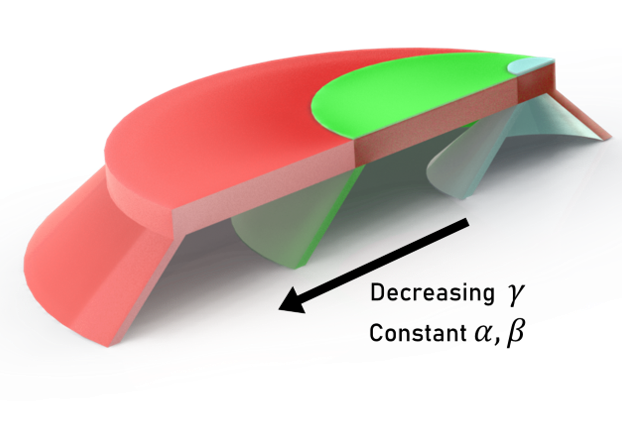}
        \centering
        \caption{}
    \end{subfigure}
    \caption{(a) Basic unit of truncated cone structure with a stub on top to distribute the force onto the inclined sidewalls. The cones can be described using Cartesian coordinates or in the $\{s,\theta, z\}$ coordinate system as shown, where $s$ is along the sidewall section, and $\theta$ is in the circumferential direction, and $z$ is in the thickness direction. (b) Non-dimensional quantities \abc{} describe the cone geometric parameters to characterize their response. Note: Independent parameters used to describe the cone are shown in black. Derived parameters are shown in blue. (c) Three cones with constant $\alpha, \beta$ but different $r_b$ show the effect of curvature, or $\gamma$, when their sidewalls are aligned. }
    \label{fig:cone_geom_params}
\end{figure}

A truncated cone is chosen as the basic unit of the architected structure due to its higher initial stiffness compared to the dome structure as described in Section \ref{se:motivation}. As seen in Figure \ref{fig:cone_geom_params}, the cone structure consists of two structural components - 1) a thin shell truncated cone with sidewall thickness \textbf{$t_w$}, height \textbf{$H$} and radius of base \textbf{$r_b$}; and 2) a stub with radius \textbf{$r_s$} and height \textbf{$h_s$}. The stub acts as a load distributor onto the edge of the cone's sidewalls. 
	
The geometric parameters of the cone are related to three non-dimensional quantities: $\alpha = \frac{t}{L}$, \textbf{$\beta = tan^{-1}(\frac{H}{r_b - r_s})$} and $\gamma = \frac{L}{r_b}$ for characterizing the mechanical behavior of the cone unit cell. $\alpha$ captures the slenderness of the sidewall, and $\beta$ captures the inclination of the cone sidewall to the horizontal. While $\alpha$ and $\beta$ would be sufficient to describe the behavior of inclined beams like in \cite{shan_multistable_2015}, another parameter is required to capture the effect of curvature of the sidewall in the case of cones presented here. This curvature can be characterized by using the radius of curvature at the base assuming all cones are tapered towards the top. This parameter is  described as $\gamma$ and non-dimensionalized by normalizing the length of inclined sidewall $L$ against the bottom radius $r_b$. As the bottom radius, $r_b$, increases for a fixed $\alpha$ and $\beta$, the curvature reduces and approaches a flat surface as pictorially depicted in the rightmost figure in Figure~\ref{fig:cone_geom_params}.

A composite sandwich structure is designed primarily to carry bending loads. Hence, the cones that form the core of such a sandwich structure locally experience both compressive and shear loads. In this study, displacement controlled loading is used to test and simulate the cone behavior. For this, a non dimensional quantity $\epsilon_v = \frac{\delta_v}{H}$ is defined for representing the vertical end-to-end shortening of the cone normalized by the cone's height during compression. Similarly, $\epsilon_h = \frac{\delta_h}{r_b-r_s}$ is defined for representing the lateral deformation normalized by the projection of the sidewall length during shear loading. 

\subsection{Printing and Fabrication of unit cells}
\label{sse:printing}
\begin{figure}
    \centering
    \includegraphics[width = \textwidth]{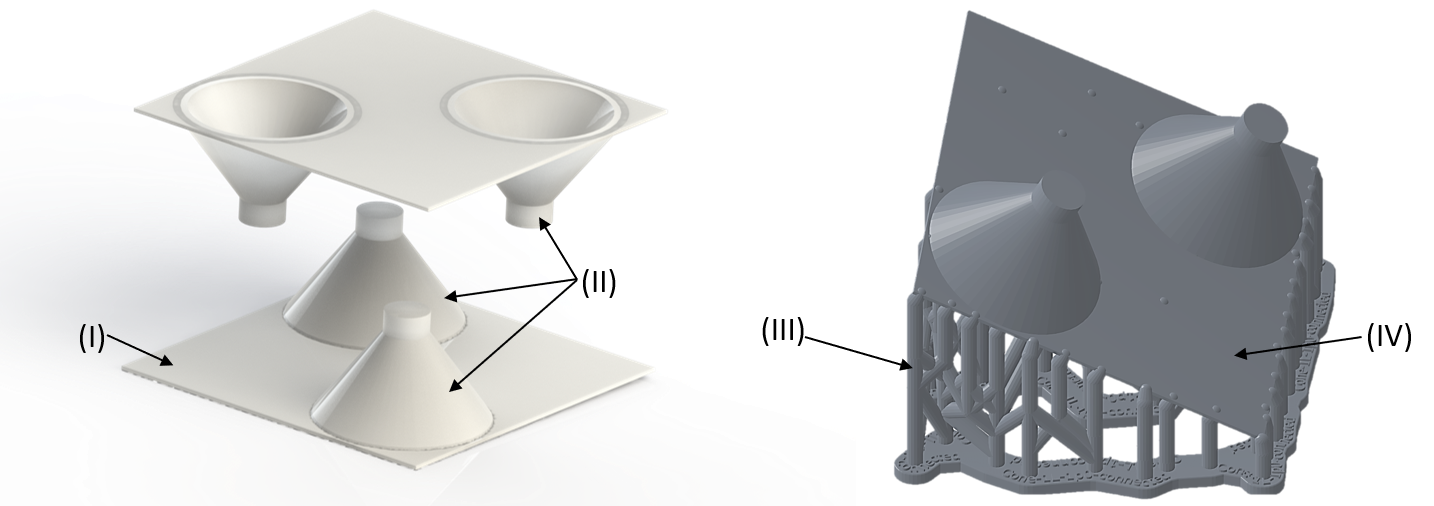}
    \caption{CAD model for printing includes (I) Membrane used for adhesively bonding core to the carbon fiber facesheet and (II) cone unit cells. Generated Formlabs PreForm file for printing with (III) support structure and (IV) final desired structure. }
    \label{fig:fabrication}
\end{figure}

The architected core structures were printed using Formlabs Form 3 SLA 3D printer. A flexible resin (manufacturer's code: FLFLGR02) from Formlabs was chosen as the material for the core structure. A layer height of 50 $\mu$m was chosen for achieving good quality prints. The printed parts consist of cones (II) and a thin membrane (I) as shown in Figure \ref{fig:fabrication}. This thin membrane serves as flat area for bonding the core to the carbon fiber facesheet. This method helped in reducing the number of support structures required during printing and also reducing the complexity of post-processing. The printed cones lie along the diagonal on a square packed grid of the thin membrane (IV) as shown in Figure~\ref{fig:fabrication} (right). These are printed on supports (III) which are removed to obtain the final part which includes I and II only. An identical part is printed and the two parts are meshed to complete a 2x2 square packed array of the truncated cones. They are glued together using Loctite 9340 as shown in Figure~\ref{fig:fabrication}. This forms one combined larger unit of the architected core structure which is then glued to carbon fiber facesheets to form a sandwich structure. Several of these units can be combined to form a core that can span a beam, plate, etc. as needed. They are also used as stacking members to form deeper beams with multiple layers of stiff carbon fiber laminates between them as shown in Figure~\ref{fig:architected_core}. 

\subsection{Material properties from experiment}
\label{sse:mat-char}
\begin{figure}[h]
     \centering
     \begin{subfigure}[b]{0.49\textwidth}
         \centering
         \includegraphics[width=\textwidth]{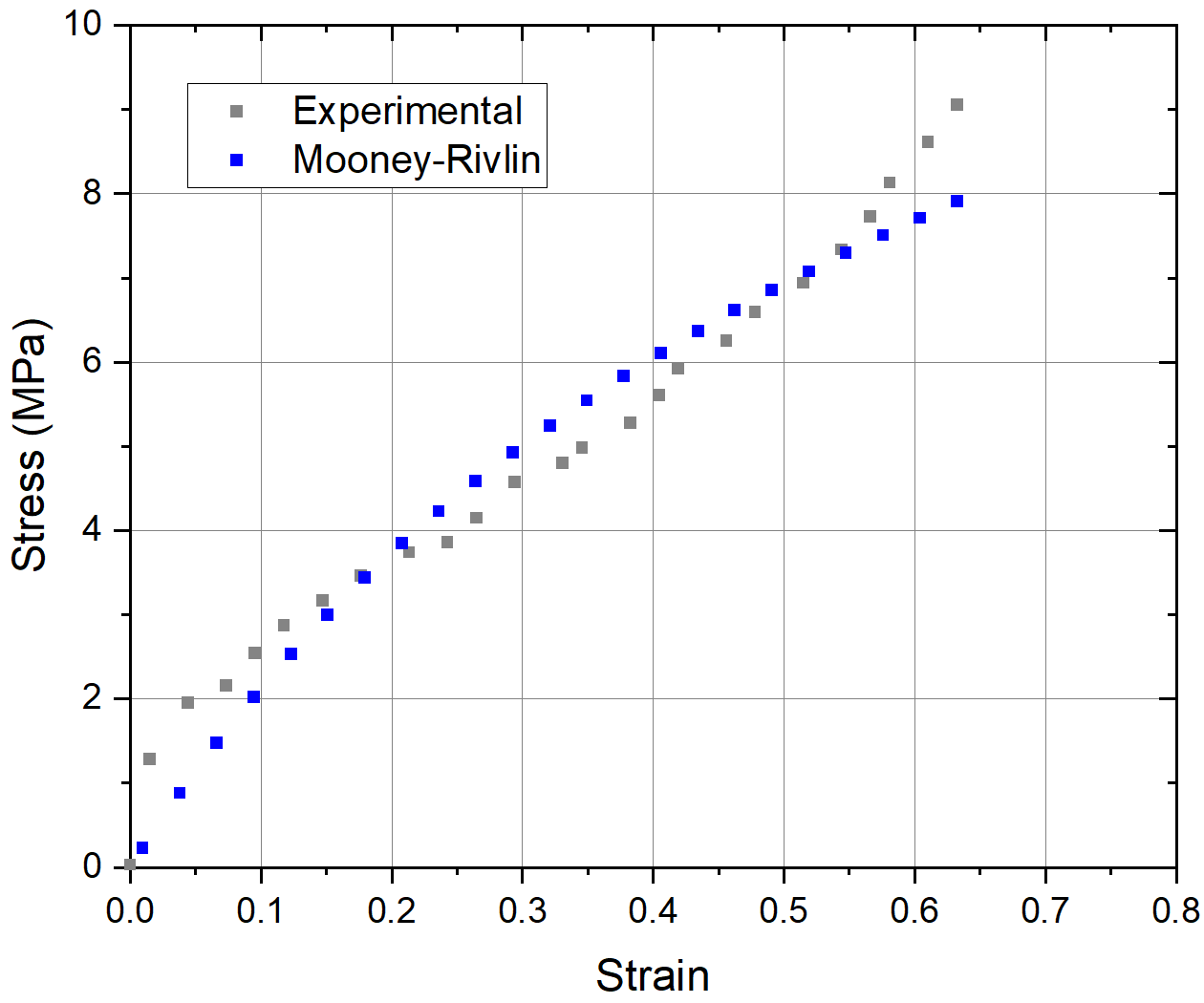}
         \caption{Stress-strain graph of dog-bone specimen conforming to ASTM D412 die-C tested to failure at 500mm/min, and evaluated as Mooney-Rivlin incompressible material for input into Abaqus.}
         \label{fig:material_property_MR}
     \end{subfigure}
     \hfill
     \begin{subfigure}[b]{0.49\textwidth}
         \centering
         \includegraphics[width=\textwidth]{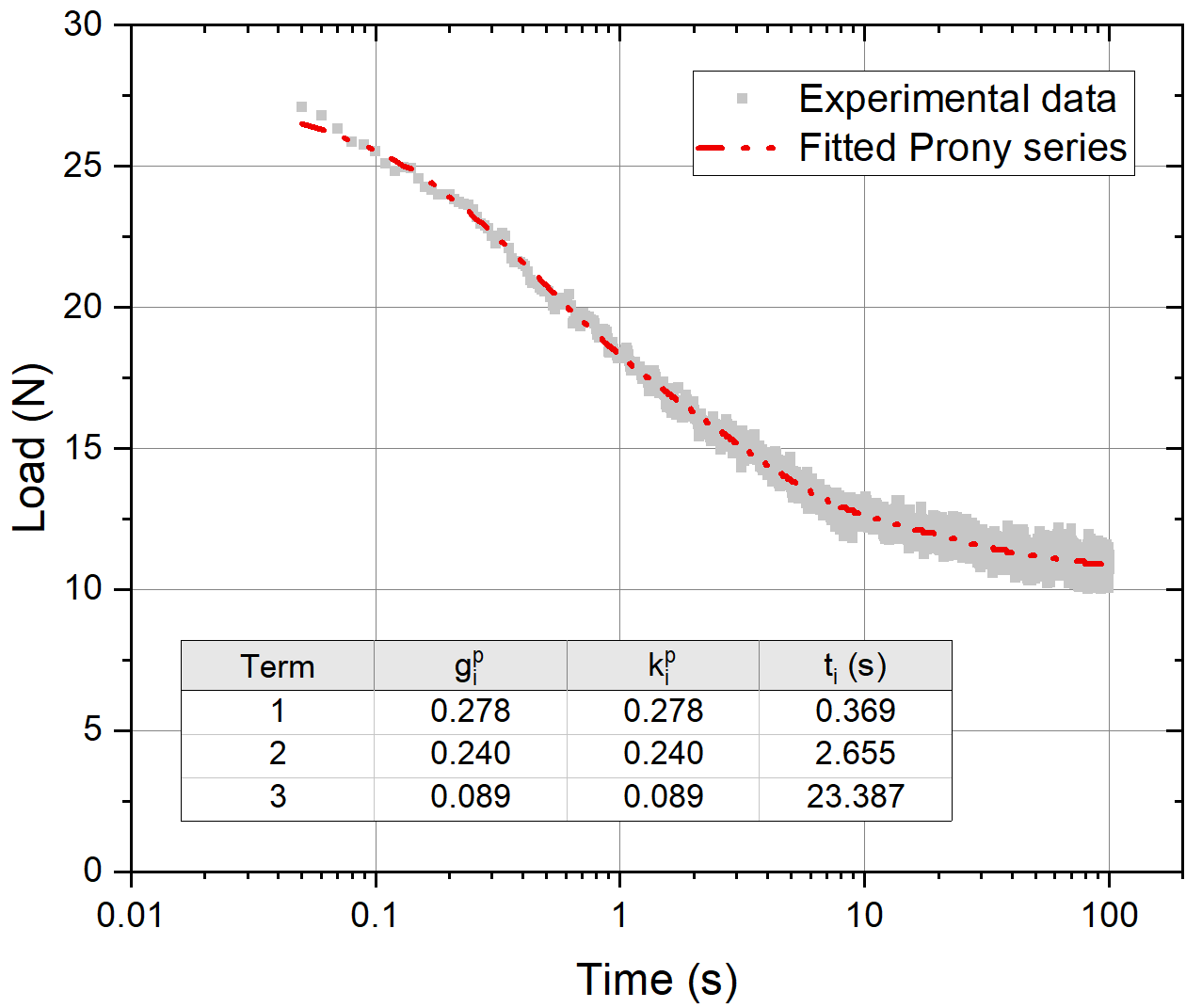}
         \caption{Time domain Prony terms fitted to experimental relaxation tests performed at constant strain. Sample conforming to ASTM D412 die-C was loaded to target strain of 5\% at a rate of 500 mm/min. }
         \label{fig:material_property_visco}
     \end{subfigure}
     \caption{Material input properties of flexible resin for finite element analysis}
\end{figure}

The flexible resin used to 3D print the unit cells was characterized to obtain input properties for use in finite element analysis discussed next in Section~\ref{sse:FEA}. For this, dog-bone specimens were printed with dimensions conforming to ASTM D412 die-C \cite{astm_d11_committee_test_nodate}. The specimens were tested at a loading rate of 500 mm/min until ultimate failure. Strain was recorded using a video extensometer. The stress-strain data was input into Abaqus and fitted to Mooney-Rivlin hyperelastic material with $\nu = 0.50$ as shown in Figure \ref{fig:material_property_MR}. The viscoelastic properties of the material were obtained using relaxation tests performed at constant strain on the dog-bone specimens as stated above. The specimens were elongated to a range of target strains from 5\%-20\% in increments of 2.5\% at 500 mm/min, and held there while the load-time response was recorded. The relaxation segment of the load-time responses were fitted to a Prony series as described in Equation \ref{eq:prony} for each of the load-time graphs, one of which is shown in Figure \ref{fig:material_property_visco}.

\begin{subequations}\label{eq:prony}
\begin{align}
    g_R(t) = 1 - \sum_{n=1}^{N} g_i^p (1-e^{-t/\tau_i}) \label{eq:prony_g} \\
    k_R(t) = 1 - \sum_{n=1}^{N} k_i^p (1-e^{-t/\tau_i}) \label{eq:prony_k}
\end{align}
\end{subequations}

\noindent where, $g_R(t)$ and $k_R(t)$ represent the shear and bulk relaxation modulus, respectively. $g_i^p$ and $k_i^p$ represent the shear and bulk relaxation parameters of the material, respectively. $\tau_i$ represent the material time constants \cite{brinkmeyer_pseudo-bistable_2012}. Similar to Brinkmeyer \cite{brinkmeyer_pseudo-bistable_2012}, $g_i^p = k_i^p$ was assumed without loss of validity. Here, a 3-term Prony series was used to fit the model the relaxation of the flexible resin material.

\subsection{Experimental setup}

\subsubsection{Unit cell compression}
\label{sse:validation-setup}

\begin{figure}[h]
    \centering
    \begin{minipage}[b]{0.53\textwidth}
        \centering
        \includegraphics[width = \textwidth]{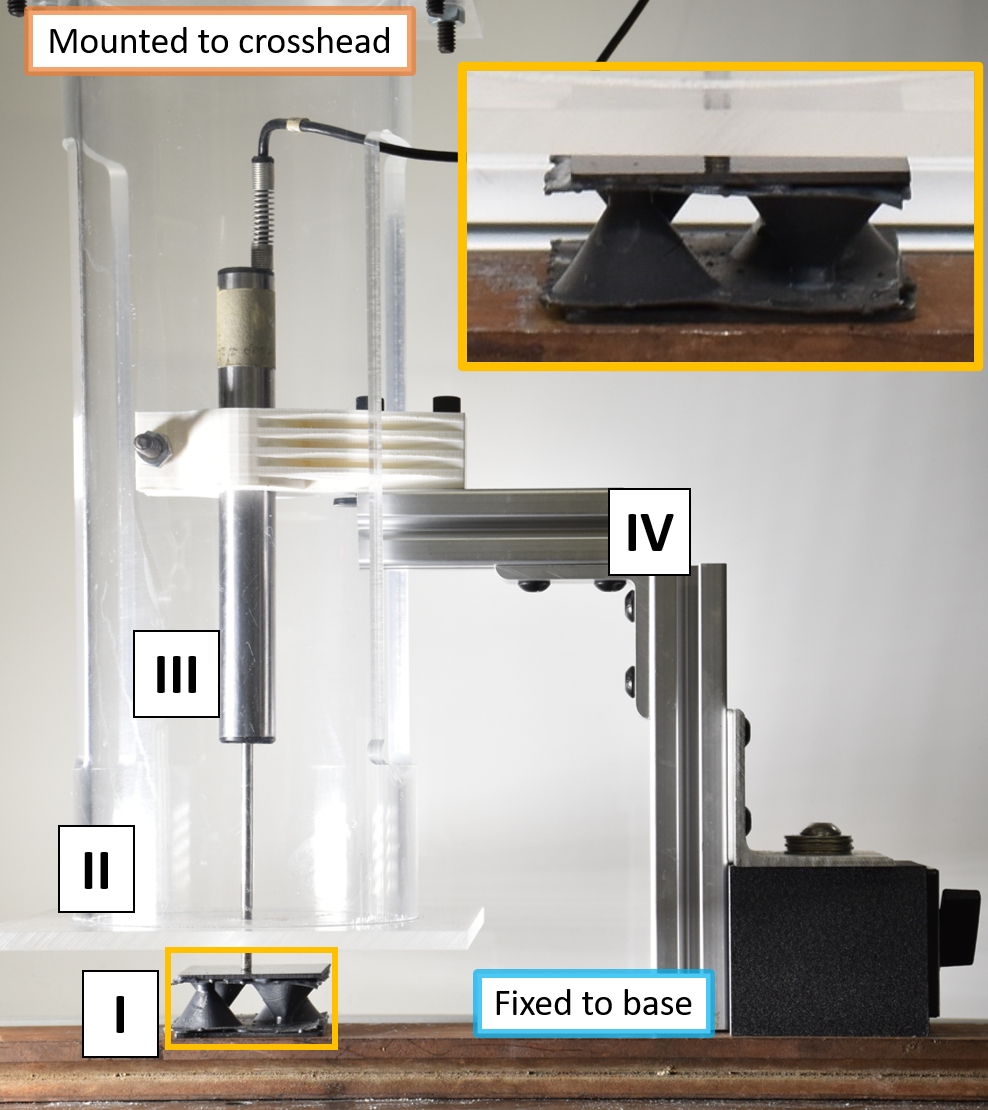}
        \subcaption[a]{}\label{fig:test_fixture}
    \end{minipage}%
    \begin{minipage}[b]{0.45\textwidth}
        \centering
        \includegraphics[width=\textwidth]{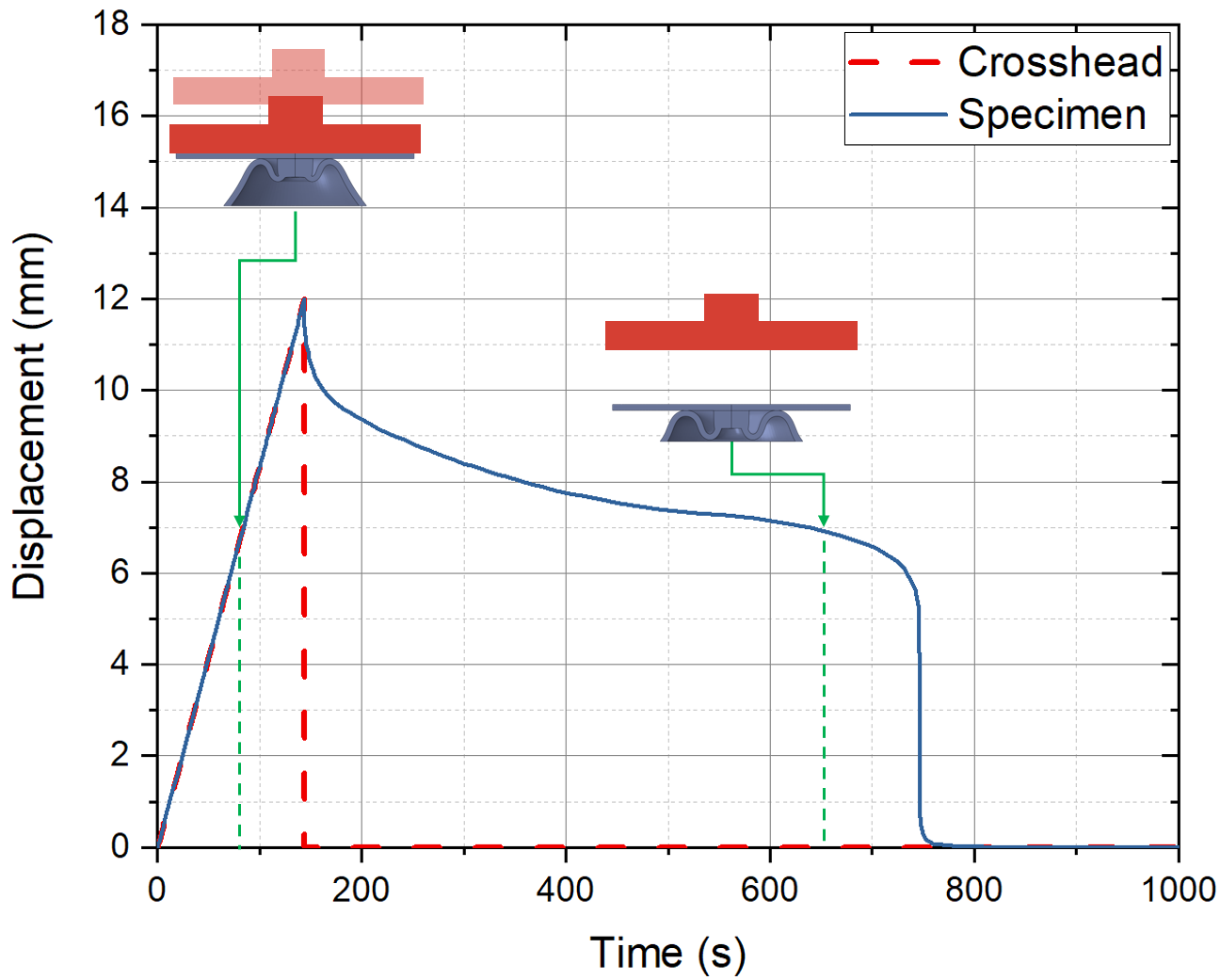}
        \subcaption[b]{}\label{fig:compressive_loading}
    \end{minipage}%
    \caption{Experimental setup for testing unit cell compression. 
    \textbf{a)} Test fixture: I) Truncated cone specimen II) Acrylic loading platen attached to the machine's crosshead. Loading platen has a pass-through hole for LVDT. III) LVDT to make contact with the specimen during both the loading and released cycle. IV) LVDT support independent of the loading platen and firmly attached to the base loading plate.
    \textbf{b)} Displacement controlled loading profile. After loading the specimen to target displacement, the platen is removed which allows the specimen to naturally recover.}
    \label{fig:expt-setup}
\end{figure}

A test fixture was fabricated to perform compression and recovery studies on the unit cells as shown in Figure \ref{fig:test_fixture}. An acrylic cylinder that is mounted to the universal testing machine's crosshead mounted load cell transfers the reaction force from the specimen during compression. An independent, rigidly mounted Linear Variable Differential Transformer (LVDT) has its probe in contact with the specimen at all times, both during compression as well as the recovery phase. An LVDT with very low spring stiffness was chosen so that the recovery phase of the specimen was not hindered by the additional load. The orange and blue lines in Figure \ref{fig:compressive_loading} represent the crosshead displacement and the displacement of the specimen respectively.

\subsubsection{Three point bend test setup}
\label{sse:3pt_bending}
\begin{figure}[H]
    \centering
    \includegraphics[width=0.95\textwidth]{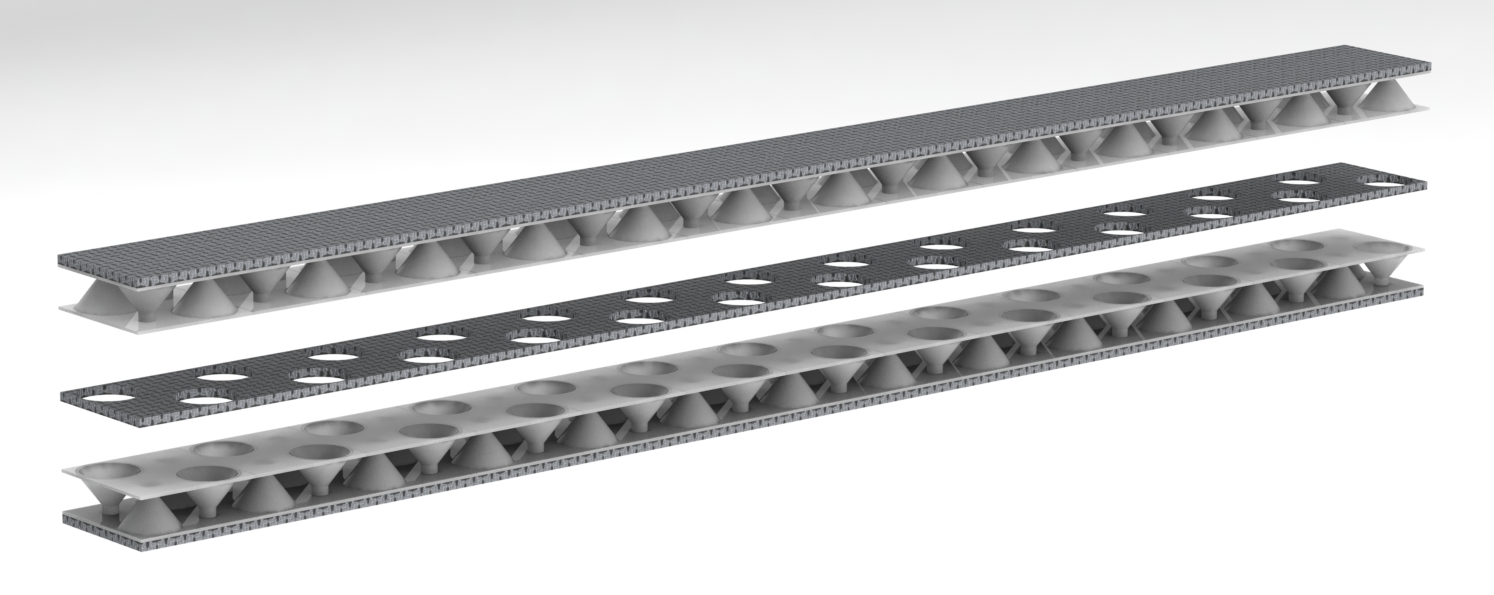}
    \caption{Basic construction of architected core sandwich structure. 3D printed core material is glued to the composite sheets to form a sandwich structure. In multi-layer composite sandwich structures, the intermediate sheets have holes that allow the core to invert itself.}
    \label{fig:3pt_bending}
\end{figure}

To verify the recoverable nature of the architected composite sandwich structure in bending, 3-point bend tests were performed. Two types of specimens were manufactured to compare the response of single-layer composite structures, and double-layered composite structures. The multi-layered structures, as shown in Figure \ref{fig:3pt_bending} have intermediate layers of carbon fiber laminate with holes to allow for unobstructed inversion of the cones under maximum compression. The holes are cut for every other cone when placed in a square packed array. Due to the alternating up-down configuration of the cones with their neighbors, the holes are required only when the cone's base lies on the intermediate layer as shown in Figure \ref{fig:3pt_bending}. The samples are glued using Loctite 9340 and cured for a minimum of 1 hour at 80\degs C. The overall dimensions of the samples, and the dimensions of the cone unit cell are shown below in Table \ref{tab:3ptDimensions}. With a carbon fiber laminate thickness of 1mm, the overall height of the samples are (15 + 2x1)=17mm for 1-layer, and (2x7 + 3x1)=17mm for 2-layer samples. The width of the samples were $4\times r_b$= 48mm and 22.4mm respectively for the 1-layer and 2-layer case. The span length in 3-point bending was maintained at 280mm. The 3-point bend tests were performed at a loading rate of 25mm/min. While the load-displacement response was obtained from the load cell, the displacement-time during recovery was obtained using video extensometer.

\begin{table}[H]

\centering
\caption{Dimensions (in mm) of unit cells in 1-layer and 2-layer samples}
\label{tab:3ptDimensions}
\begin{tabular}{l c c c c c c c c c c}
\toprule

\multirow{2}{*}{} &
  \multirow{2}{*}{$\boldsymbol{H+h_s}$} &
  \multirow{2}{*}{$\boldsymbol{h_s}$} &
  \multirow{2}{*}{$\boldsymbol{H}$} &
  \multirow{2}{*}{$\boldsymbol{r_b}$} &
  \multirow{2}{*}{$\boldsymbol{r_s}$} &
  \multicolumn{3}{c}{\multirow{2}{*}{$\boldsymbol{t_w}$}} &
  \multicolumn{2}{c}{Total sample dimensions} 
  \\ \cline{10-11} 
    &    &     &     &     &     & \multicolumn{3}{c }{} & Height & Width \\ \midrule
\multicolumn{1}{c }{\textbf{1 Layer}} & 15 & 3   & 12  & 12  & 3   & 0.5   & 1     & 1.5   & 17     & 48    \\ \midrule
\multicolumn{1}{c }{\textbf{2 Layer}} & 7  & 1.4 & 5.6 & 5.6 & 1.4 & 0.23  & 0.47  & 0.70  & 17     & 22.4  \\ \bottomrule
\end{tabular}

\end{table}

\subsection{Exploring parametric design space using FEA}
\label{sse:FEA}
\begin{figure}[h]
    \centering
    \includegraphics[width = \textwidth]{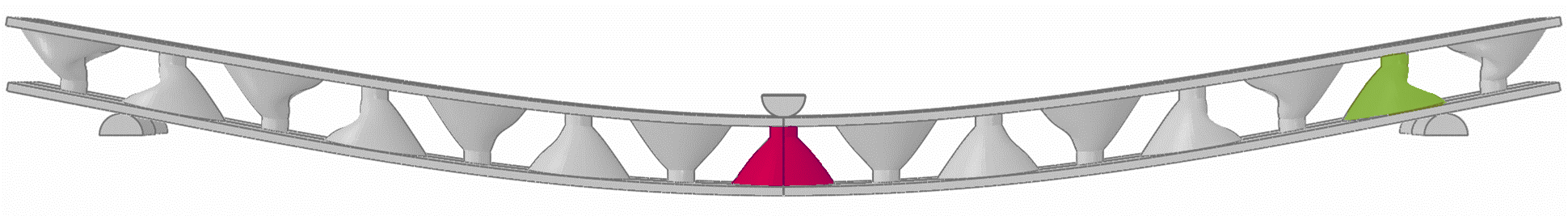}
    \caption{Cones experience combination of shear and compressive loading when the sandwich structure is subjected to flexure. Cones near the center (red) experience axial loading, while those away from the mid-span (green) experience a combination of axial and shear loading.}
    \label{fig:inside_beam}
\end{figure}

Finite element analysis of the cones was performed to computationally explore the design space under compressive and shear loading. The simulations were performed using the commercial finite element analysis software package, Abaqus. The unit cells, when part of the architected core structure undergo both shear and compression as shown in Figure \ref{fig:inside_beam}. Hence, it is important to characterize their behavior under a combination of compression and shear loading. All simulations were performed by loading the facesheet  attached to the cone. Displacement controlled loading was used to compress and shear the cones to their target values. Upon reaching the target displacement value, the boundary condition was released allowing the cones to relax and unload. The simulations were performed as implicit dynamic analyses, and the sudden change in boundary condition that would normally induce high frequency vibrations was damped through additional numerical damping that was introduced by performing time stepping using the backward Euler method.

Input material properties for the woven carbon fiber facesheet were chosen based on typical values as $E$=68 GPa and $\nu$=0.2 \cite{MitsubishiRayonCarbonFiber&Composites2014}. The density  of the carbon fiber facesheet was 1.85g/cm$^3$. The material properties for the flexible resin were obtained from tests described in section \ref{sse:mat-char} and shown in figures \ref{fig:material_property_MR} and \ref{fig:material_property_visco}. Density of the resin after 3D printing and curing was measure to be 1.15 g/cm$^3$

\subsubsection{Combined compression and shear loading}
\begin{figure}[h]
    \centering
    \includegraphics[width=0.85\textwidth]{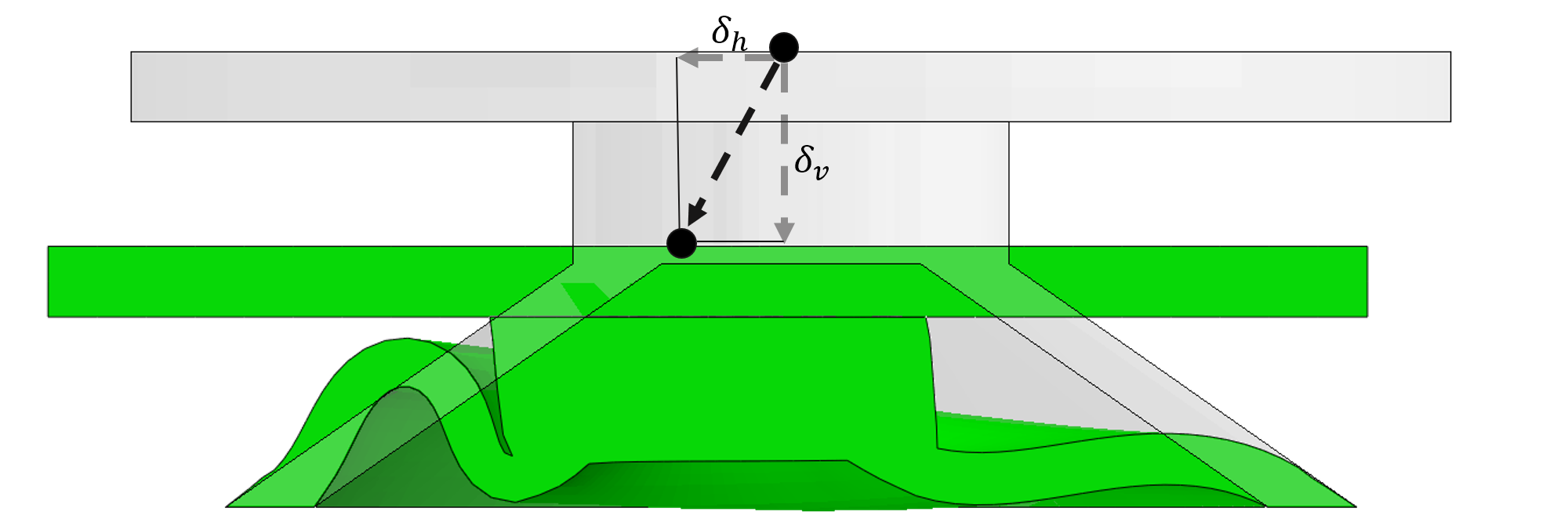}
    \caption{Combined loading with compressive and shear loading vectors.}
    \label{fig:combi-loading}
\end{figure}

To simulate the combined shear and compressive response of the unit cells, a displacement controlled method was used in Abaqus. A kinematic coupling constraint was applied to a control point as shown in Figure \ref{fig:combi-loading}. When the lateral displacement, $\delta_h = 0$, the problem reduces to an axisymmetric problem which reduces computational time. This loading scenario occurs under pure compression, or directly under the loading roller in a 3-point bend beam specimen as shown in red in Figure \ref{fig:inside_beam}. In Abaqus, CAX4H - a 4-noded bilinear axisymmetric quadrilateral, hybrid element was chosen to mesh the model due to  incompressibility of the hyperelastic material. The axisymmetric section of the structure was meshed with the minimum edge length of the element to be $\sim$0.1mm to ensure convergence of the mesh, and also optimize for computation time. 

When simulating the combined loading case incorporating both shear and compression, i.e. when $\delta_h \neq 0$, the axial symmetry of the problem no longer exists. However, a plane of symmetry along the shear loading direction reduces the problem to half symmetry. In Abaqus, C3D8RH - an 8-noded linear brick with reduced integration and hybrid formulation was chosen to mesh the model using the medial axis meshing algorithm. Element edge length of $\mathtt{\sim}0.5$mm was used to satisfy convergence conditions. 

The simulation was split into two steps: \begin{enumerate}
    \item In the first simulation step, a displacement controlled loading was performed at a constant rate of 5mm/min or 12s/mm to simulate quasi-static loading. The target displacement $\delta = \sqrt{\delta_v^2 + \delta_h^2}$ was used to determine the total loading step time as $\delta \times 12s/mm$.
    \item Upon reaching the target displacement, the enforced boundary condition was released allowing the viscoelastic material undergo relaxation as shown in Figure \ref{fig:compressive_loading}
\end{enumerate} 
The reaction force and displacement were recorded throughout the simulation at the control point. Analyzing the vertical displacement of the control point after unloading the cones allows us to classify the stability of the cones as monostable, bistable, or pseudo-bistable.

\subsubsection{Scaling of unit cells}
\label{ssse:scalability}

The architected core structure must retain its characteristic load response and recovery profile when scaled in size to verify that \abc{} are sufficient to characterize the unit cell behavior. To verify this, three cones were chosen for compression analysis - a standard cone with $r_b = 12$mm, $r_t = 3$mm, $H = 12$mm, $t_w = 1$mm, $r_s = 3mm$ and $h_s = 3mm$, and two other scaled models with $r_b, r_t, H, t_w$, $r_s$ and $h_s$ scaled as $\frac{1}{2}$x and 2x their original values as described in Table \ref{table:scaling-setup}.  The cones were compressed to a total displacement of $(H+h_s)$ at a constant loading rate of 5mm/min as this was a standard loading rate used in validation tests. 

\begin{table}[H]
\centering
\caption{Scaled dimensions (in mm) of cones used in the scaling stability tests.}
\label{table:scaling-setup}
\begin{tabular}{c|c|c|c} 

\toprule
\textbf{Geometric Parameter} & {$1/2$x scale}  & {$1$x scale} & {$2$x scale}  \\ 
\midrule
$r_b$ (mm) & 6      & 12  & 24   \\ 
$r_t$ (mm) & 1.5  & 3   & 6    \\ 
$t_w$ (mm) & 0.5  & 1   & 2    \\ 
$H$ (mm)    & 6      & 12  & 24   \\ 
$r_s$ (mm) & 1.5  & 3   & 6    \\ 
$h_s$ (mm) & 3       & 3   & 3    \\ 
$\delta_v$ & 9     & 15  & 27   \\
\bottomrule
\end{tabular}

\end{table}

\section{Results and discussion} \label{se:results}

\subsection{Validation of single unit cell compression simulation}
\label{sse:validation}
\begin{figure}[H]
\centering
\begin{minipage}{0.49\textwidth}
    \centering
    \includegraphics[width=\textwidth]{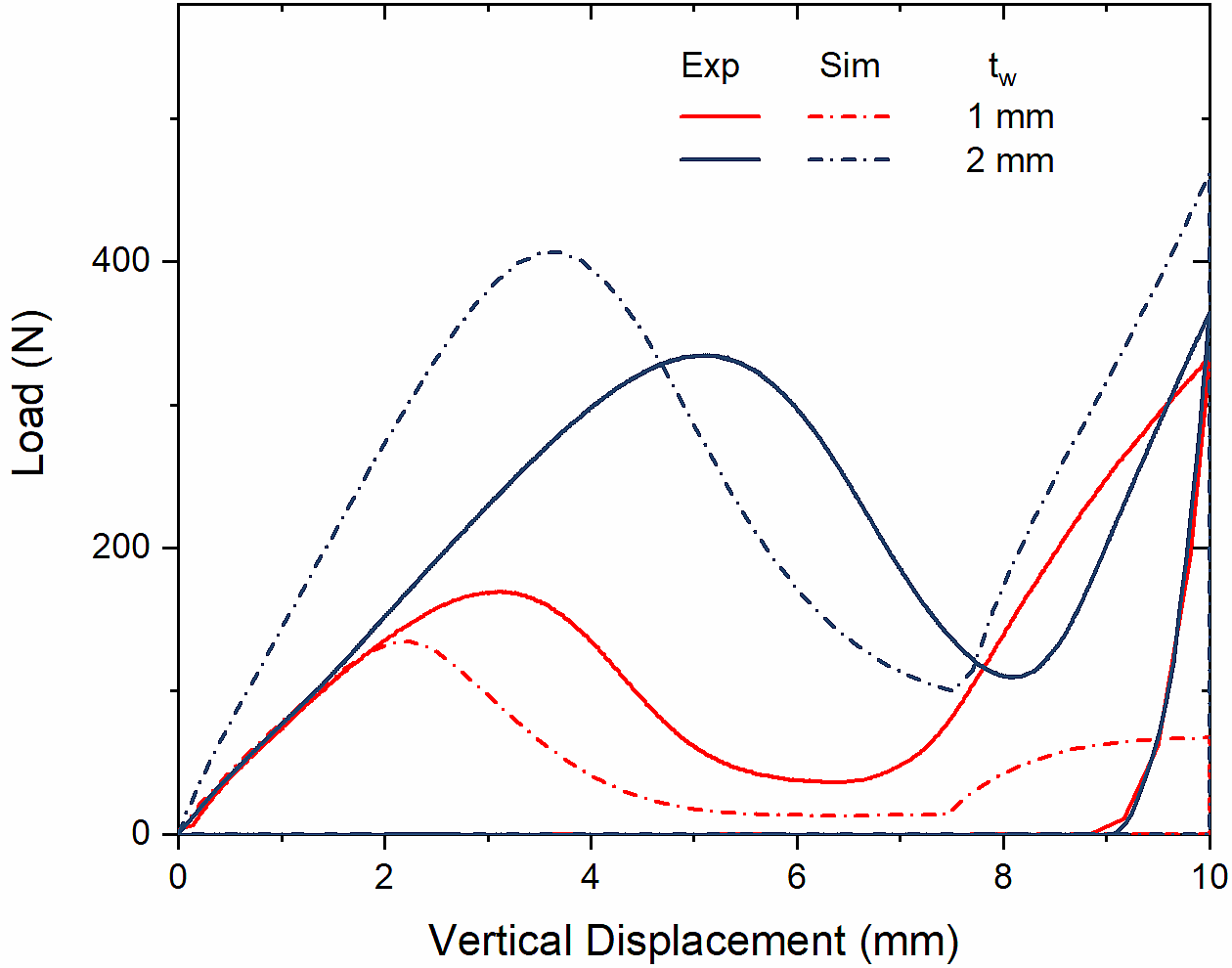}
    \subcaption[a]{Load-Displacement validation}\label{fig:exp_val_load_disp}
\end{minipage}%
\hfill
\begin{minipage}{0.49\textwidth}
    \centering
    \includegraphics[width=\textwidth]{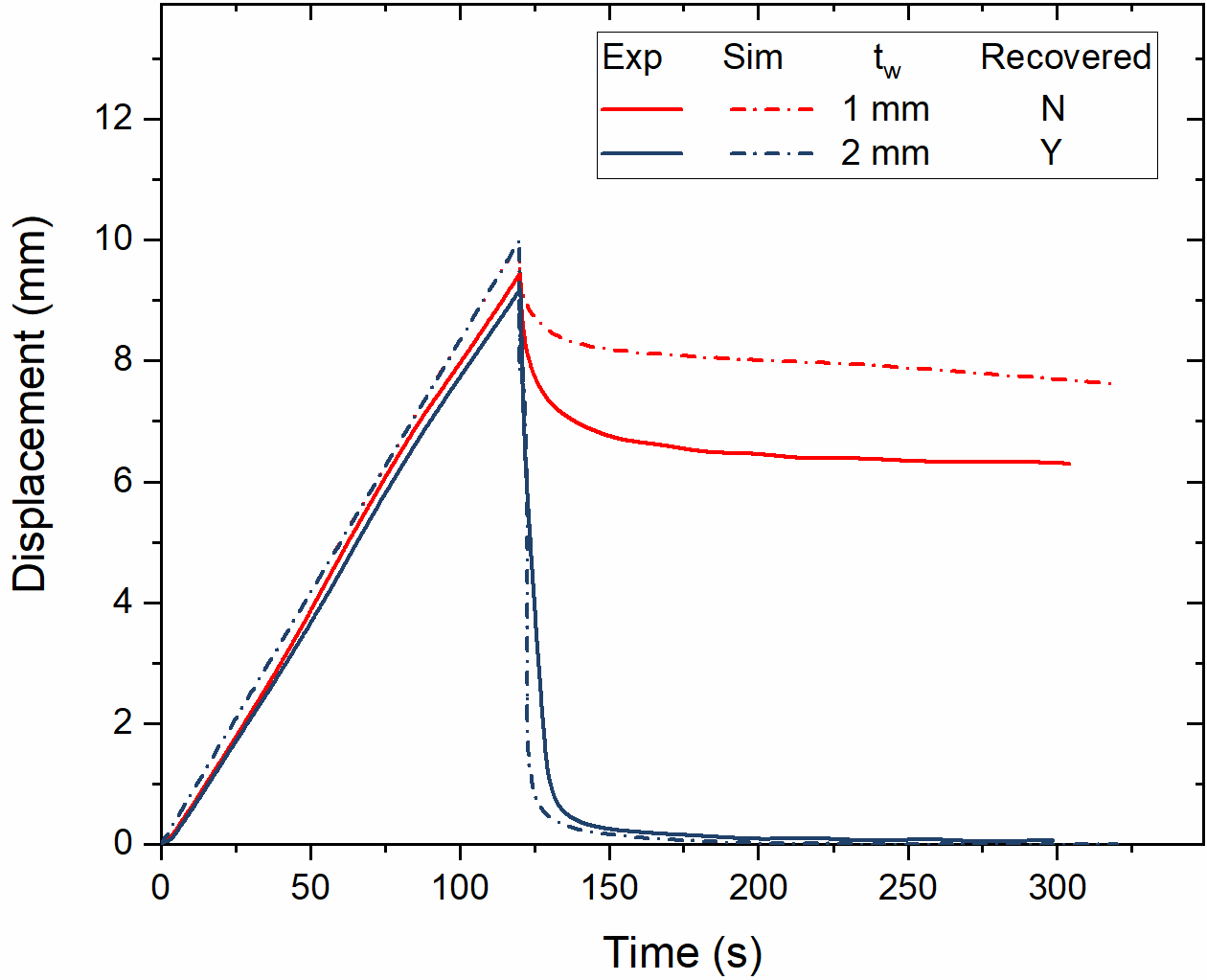}
    \subcaption[b]{Displacement-Time validation}\label{fig:exp_val_disp_time}
\end{minipage}%
\caption{Experimental validation of simulation results} \label{fig:exp_val}
\end{figure}

Prior to analyzing the behavior of architected core structures with repeating units, the finite element analysis results were validated at the unit cell level. The results were validated against compression tests performed with the setup described in section \ref{sse:validation-setup}. As the material inputs were obtained from experimental tests, the effects of the cone geometric properties were tested by direct compression tests on the unit cell. For this, two cones with $r_b=12$mm, $r_s=3$mm, $H=12$mm, $h_s = 3$mm and wall thicknesses of 1mm and 2mm, respectively, were used. The simulation results of load-displacement (Figure \ref{fig:exp_val_load_disp}) and displacement-time (Figure \ref{fig:exp_val_disp_time}) were in good agreement with the experimental results, thus validating the simulation for further studying the stability conditions for a range of geometric parameters.

\begin{figure}[H]
    \centering
    \includegraphics[width=\textwidth]{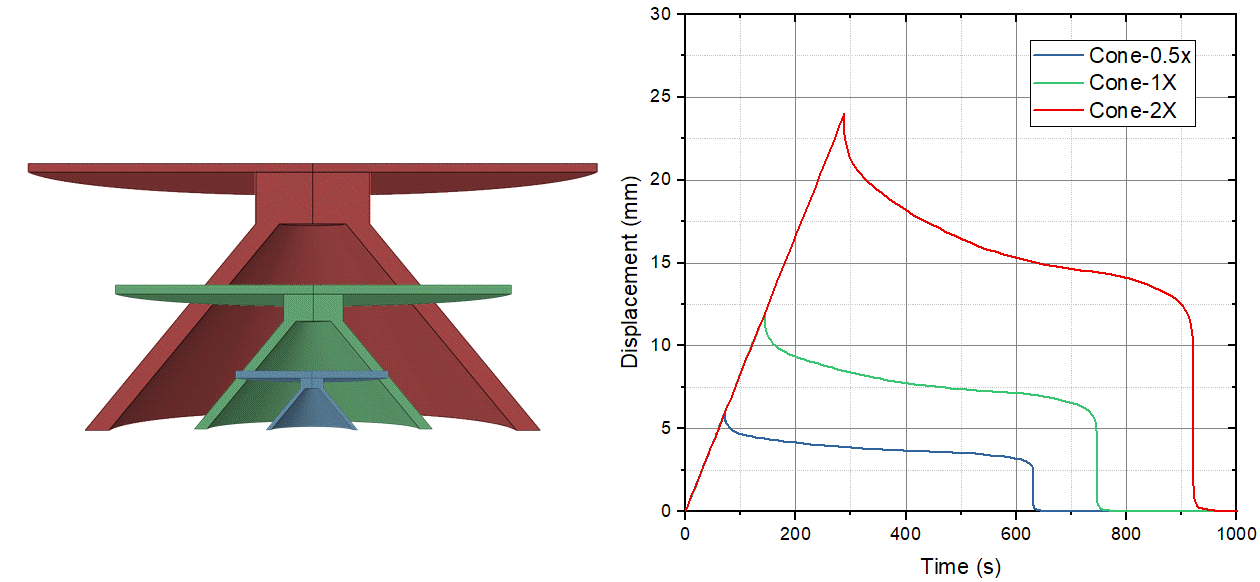}
    \caption{Scaling analysis test results show identical stability behavior for cones with same $\alpha, \beta, \gamma$ values, but scaled as 0.5x, 1x, and 2x with bottom radii of 6mm, 12mm, and 24mm respectively.}
    \label{fig:scaling-test-results}
\end{figure}

Along with validation of the simulation against experimental data, a scaling analysis was performed to verify that the non-dimensional parameters \abc{} used to characterize a unit cell are not affected by the length scale of the cones. Three cones with identical \abc{} were scaled by their height and analyzed in compression as described in section \ref{ssse:scalability}, and the results are shown in Figure \ref{fig:scaling-test-results}. The overall stability of the three cones remained the same, that is, pseudo-bistable for this combination of \abc{}. Further, the time taken by the unit cells to return to their original configuration was within a small range, showing that the non-dimensional parameters are sufficient to describe the behavior of these cones. It must be noted that this has been verified only for unit cells with their thicknesses in the order of millimeters. At a  smaller length scale, the behavior may not be sufficiently captured owing to other phenomena occurring within the polymeric resin material.

\subsection{Parameter space for combined shear and compression loading}
\label{sse:results_parametric}
The parametric space of the unit cell was evaluated by considering the response of the cones in axial compression as well as in combined compression and shear loading. Particularly, three criteria were used for evaluation - critical buckling load, displacement to achieve buckling, and the post-buckling stability. 

\subsubsection{Buckling load and displacement}
\begin{figure}[h]
    \centering
    \begin{subfigure}[b]{0.4\textwidth}
        \centering
        \includegraphics[width=\textwidth]{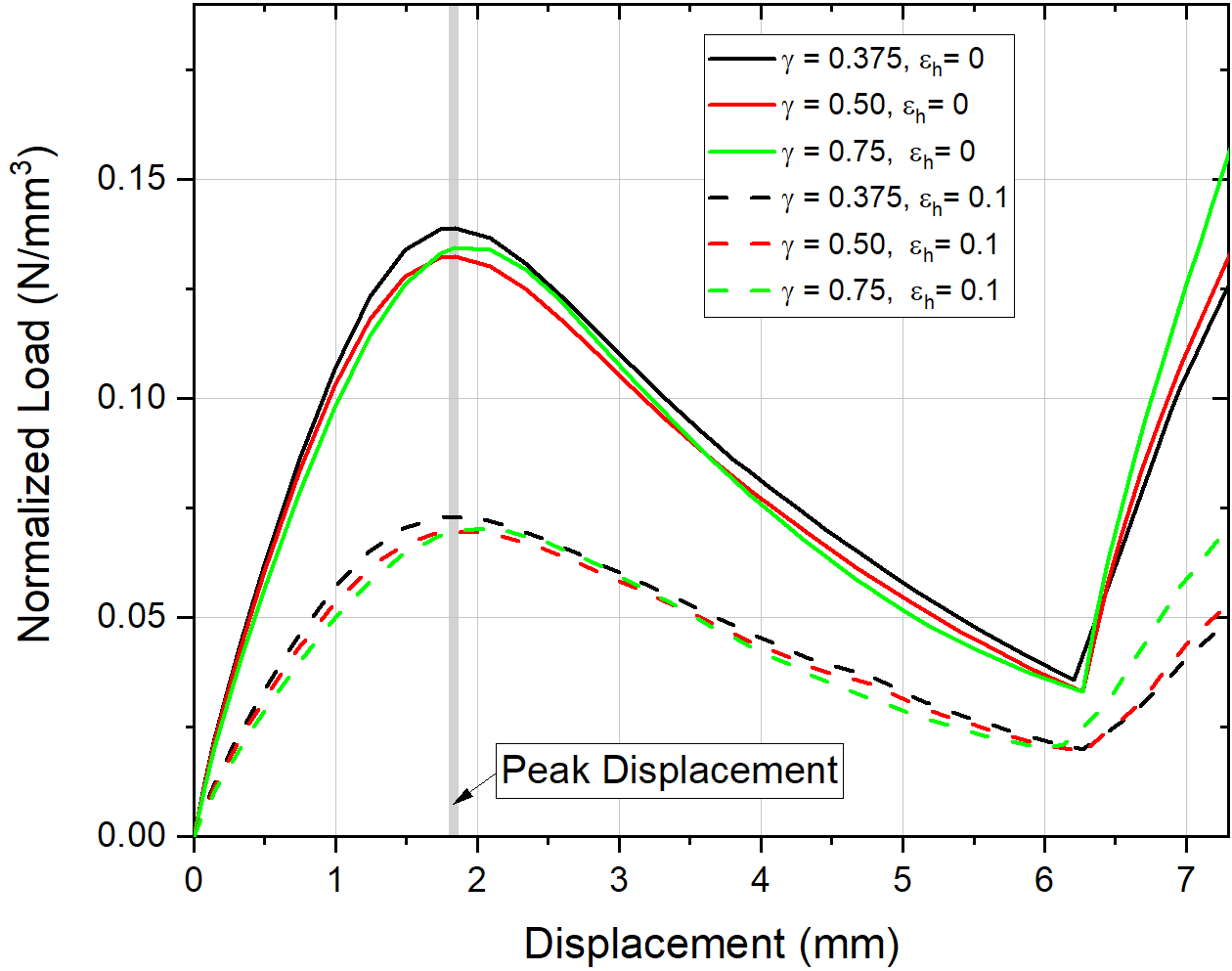}
        \caption{}
        \label{fig:scaled_LD}
    \end{subfigure}
    \vfill
    \begin{subfigure}[b]{0.4\textwidth}
        \centering
        \includegraphics[width=\textwidth]{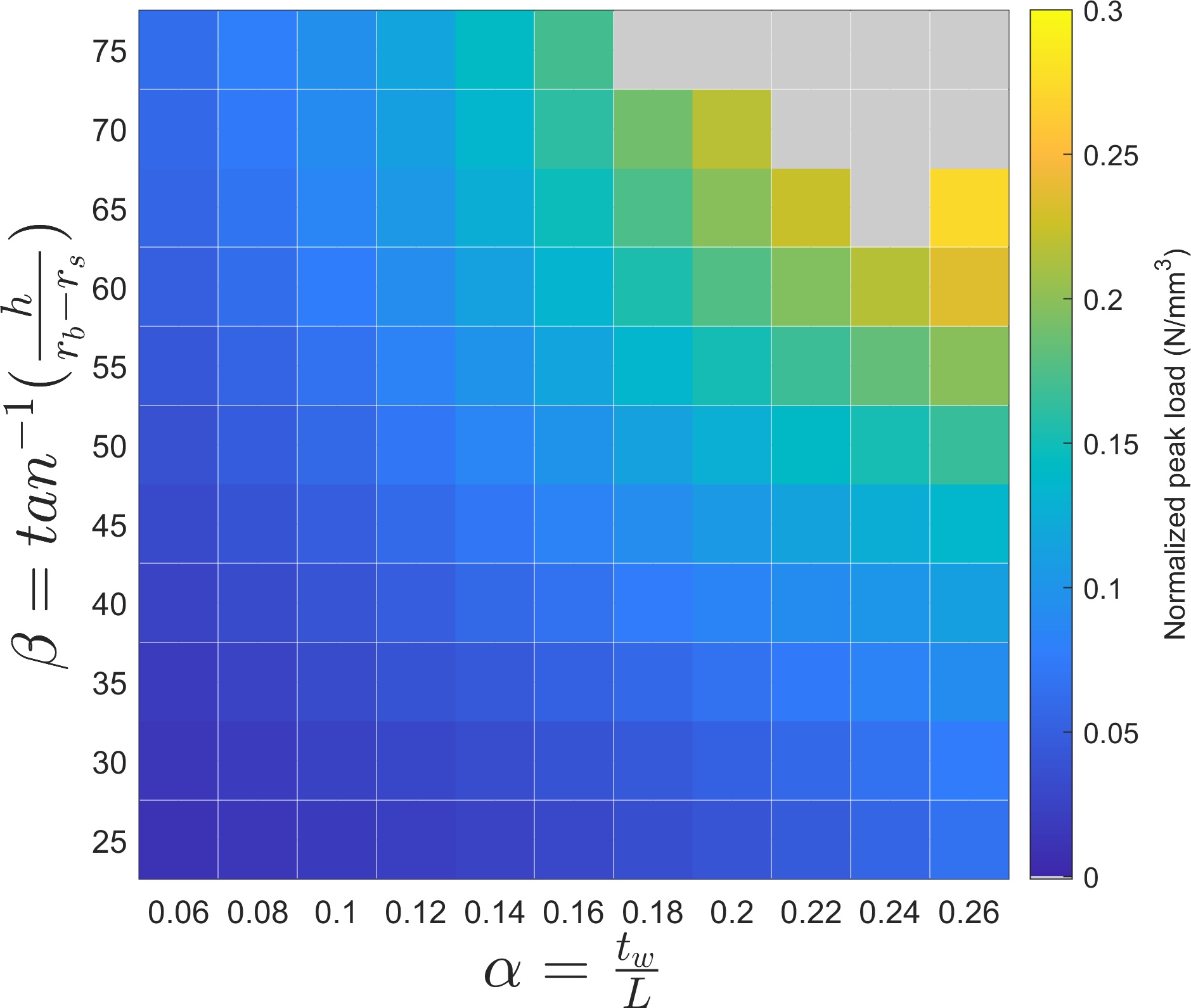}
        \caption{}\label{fig:peak_load}
    \end{subfigure}
    \hfill
    \begin{subfigure}[b]{0.4\textwidth}
        \centering
        \includegraphics[width=\textwidth]{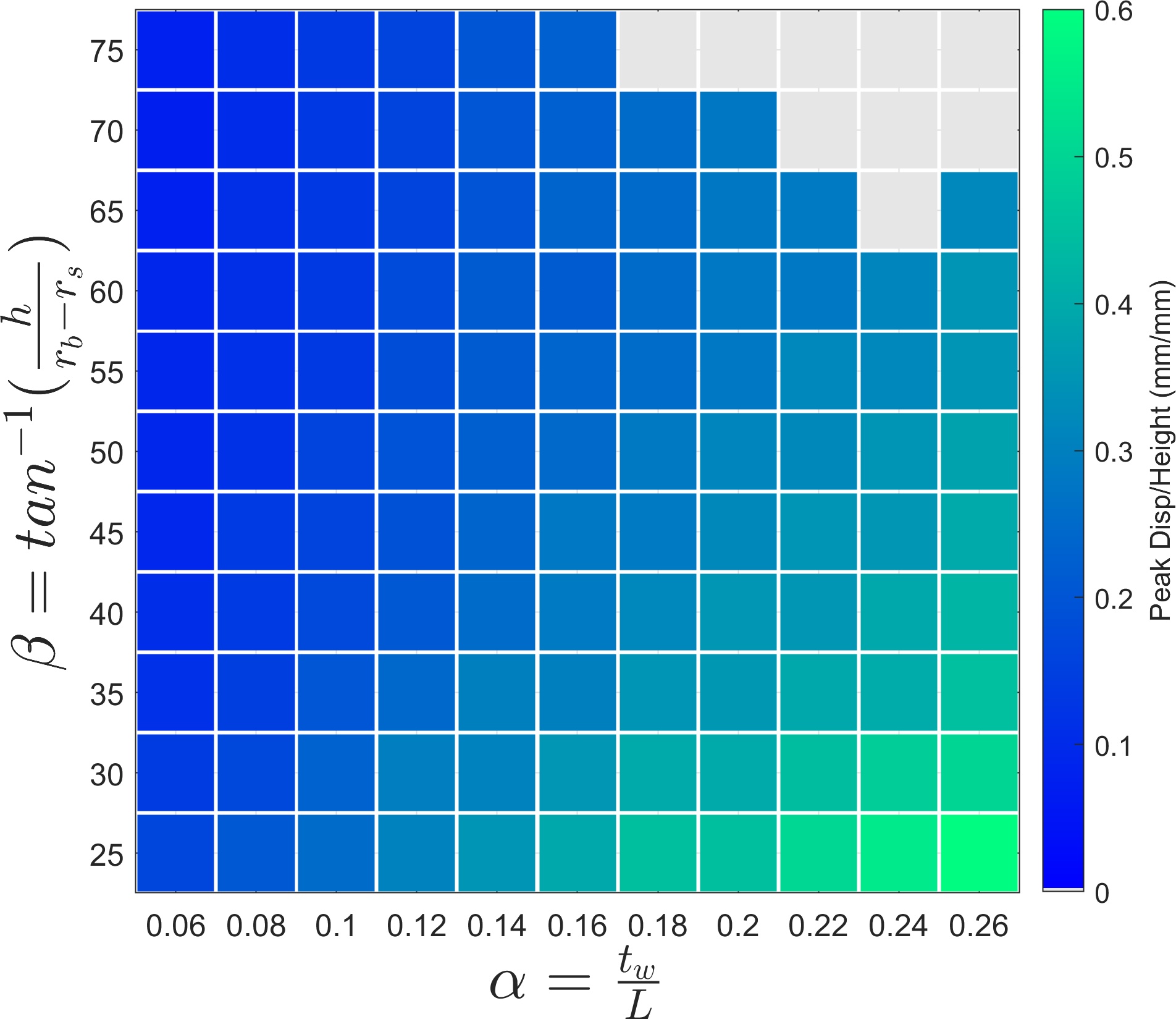}
        \caption{}\label{fig:peak_disp}
    \end{subfigure}
    \caption{(a) Peak buckling load scaled by the volume of the curved sidewalls of the cone. Here, $\alpha=0.18, \beta=55$\degs and $\epsilon_v = 1$ (b) Peak buckling load normalized by the printed volume of curved sidewalls of the conic frusta. (c) Peak displacement normalized by height of the cones.}
\end{figure}

The critical buckling loads for unit cells were obtained by finding the first peak in the load-displacement response graph. The vertical displacement at which this peak buckling load occurs is also noted as it is of interest for a designer. Varying the $\gamma$ while keeping $\alpha,\beta$ constant, we see that the peak loads when normalized by the volume of curved shell section (Curved Surface Area $\times$ Thickness =  $\pi(r_b+r_s)\sqrt{H^2 + (r_b-r_s)^2} \times t_w$)  of the unit cell are close to each other as seen in Figure \ref{fig:scaled_LD}. The volume of the stub is ignored as it does not contribute to the axial load until secondary contact occurs (see figure \ref{fig:initial_cone_vs_domes})

Similarly, the peak displacement to achieve buckling are also very close to each other, at $\sim$1.85mm, or $\epsilon_v\sim 0.245$. This shows that the peak buckling load and peak displacement to buckling are dependent on the parameters $\alpha, \beta$ and not dependent on $\gamma$, or indirectly, on the printed volume of the sidewalls of the unit cell. When a perturbation is added in the form of lateral displacement, we see the peak buckling loads drop (dashed lines in Figure \ref{fig:scaled_LD}, but the peak displacements to buckling remain the same.

Figure \ref{fig:peak_load} shows the parametric space for $\alpha, \beta$ with a constant $\gamma$ of 0.75. We see that the normalized peak load is directly proportional to both $\alpha$ and $\beta$. This is due to the high stiffness provided by thicker walls that is captured in $\alpha$ and a more vertical structure resulting from a larger $\beta$. Smaller $\beta$ produces a shallow cone structure that is susceptible to buckling, while a taller, more upright cone produced by a larger $\beta$ behaves more like a cylinder.

The peak displacement to buckle is directly proportional to $\alpha$ and inversely proportional to $\beta$. A thicker sidewall (larger $\alpha$) is stiffer and provides flexural  rigidity delaying the buckling of the cones. However, for a shallow cone (smaller $\beta$), the applied energy is more directly used for bending rather than for axial compression, making it susceptible to buckle early.

\subsubsection{Post Buckling Stability}

After reaching the critical buckling load, the unit cells have a negative stiffness region followed by stiffening due to secondary contact (Figure \ref{fig:initial_cone_vs_domes}). When the displacement boundary condition is released, the cones display 1 of these 3 possible behaviors - \begin{enumerate}
    \item The cones immediately recover to their initial unloaded state, exhibiting monostable behavior. This occurs when the elastic strain energy stored in the system is sufficient to restore the structure to its original original configuration.  
    \item The cones get locked in their deformed configuration, exhibiting bistable behavior. This occurs when the stored stored strain energy in the system is insufficient to cross the energy barrier required to return to its original configuration. 
    \item The cones exhibit a pseudo-bistable behavior, where for a certain period of time, they appear to be locked in the deformed configuration due to an energy barrier that cannot be overcome with the stored strain energy. However, as the material dissipates energy through viscoelastic relaxation, the energy barrier that is `locking' the structure in its deformed configuration reduces until the point when the elastic energy is sufficient to restore the structure to its original configuration.  
\end{enumerate}

Table \ref{tab:stability} shows the stability conditions for the unit cells with varying parameters. Purely compressive loading cases (axisymmetric), $\epsilon_h=0$, are shown in the first column with each row representing a different value of $\gamma$. For $\gamma = 0.75$, we see that cones with slender side walls identified with small $\alpha$, and large angle $\beta$ (tending towards vertical side walls) exhibit bistable behavior. With increasing $\alpha$ and decreasing $\beta$, the structures tend towards monostable behavior - shown in green. In the transition zone between bistable and monostable behavior, the structures exhibit pseudo-bistable behavior represented by blue. This behavior is attributed to the viscoelastic dissipation in the material. In an elastic material without viscoelastic dissipation, this blue transition region of pseudo-bistability would not exist. 

With decreasing $\gamma$ values, we see a tendency of the structures to return to their initial state. Although the volume normalized buckling loads of the structures are similar to each other with varying $\gamma$ and constant $\alpha, \beta$ as seen above in Figure \ref{fig:scaled_LD}, their post-buckling stability conditions are different. This difference is attributed to the distribution of axial and bending energy in the curved sidewalls of the cones, with the axial direction being along the length $L$. With a smaller radius of curvature (larger $\gamma$), the restoring moment required to return the structure to its original configuration is very small compared to the case with smaller $\gamma$. This can be seen by comparing the deformed shapes of two cones with identical $(\alpha, \beta)$ = (0.18, 55\degs) and $\gamma =$ 0.75 and 0.375, respectively. From Table \ref{tab:stability}, we see that these cones exhibit pseudo-bistability and monostablity, respectively. By looking at the deformed shapes at $\epsilon_v = 1$, we see that the curvature of the cones, $\kappa_\theta$, influence the bending curvature of the sidewalls, i.e. $\kappa_s$ (refer to Figure \ref{fig:cone_geom_params}). With a smaller radius of curvature, the deformed sections of the sidewall rotate inwards to have a `locking' effect to keep the stub in the lower position. Without viscoelastic dissipation, this locking effect will lead to bistability. However, in a viscoelastic material, we observe this effect diminish over time which eventually results in the cone returning to its undeformed state. With larger $\gamma$, or smaller radius of curvature, we again see this `locking' effect, making the cones less likely to return to their undeformed configuration, exhibiting bistability.

Perturbations, such as lateral loading, nudge the structures towards recovering from their deformed configurations. Monostable structures retain their monostability. Pseudobistable structures with \abc\ close to monostability on the parametric map are nudged towards monostability. Similarly, unit cells exhibiting bistability in the non-perturbed ($\epsilon_v = 0$ case) but close to pseudo-bistability in the parametric maps are nudged towards pseudo-bistability in the perturbed case. Increasing the perturbation, or the shear loading in this case, further pushes the structures towards monostability. Perturbed unit cells are largely monostable or pseudo-bistable case because of the restoring tensile force that is created on the section of sidewalls on the opposite side of loading direction. This stretching that occurs on the far side of the conic shells is responsible for restoring the structure to its original undeformed configuration.

\begin{table}[H]
    \centering
    \begin{tabular}{>{\centering\bfseries}m{0.3in}  >{\centering}m{2.8in} >{\centering\arraybackslash}m{2.8in}}
        \toprule
        $\boldsymbol{\gamma}$ & $\boldsymbol{\epsilon_v = 1, \epsilon_h = 0}$ & $\boldsymbol{\epsilon_v = 1, \epsilon_h = 0.1}$ \\
        \midrule
        0.75 & 
        \includegraphics[width=0.4\textwidth]{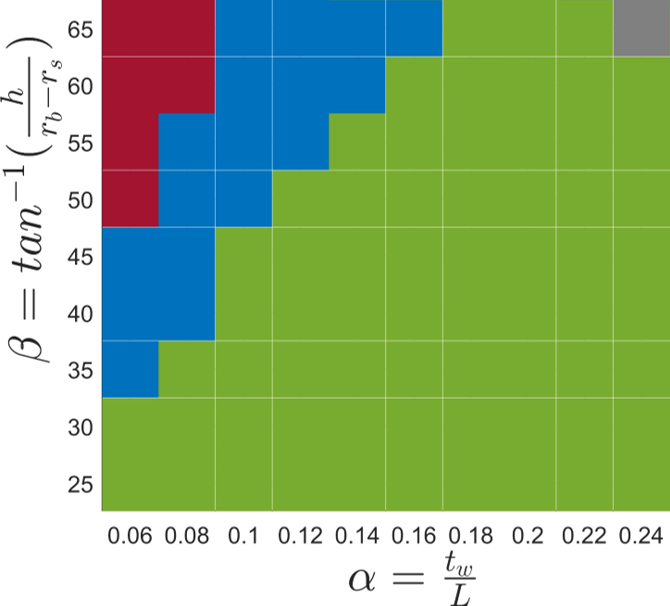} & 
        \includegraphics[width=0.4\textwidth]{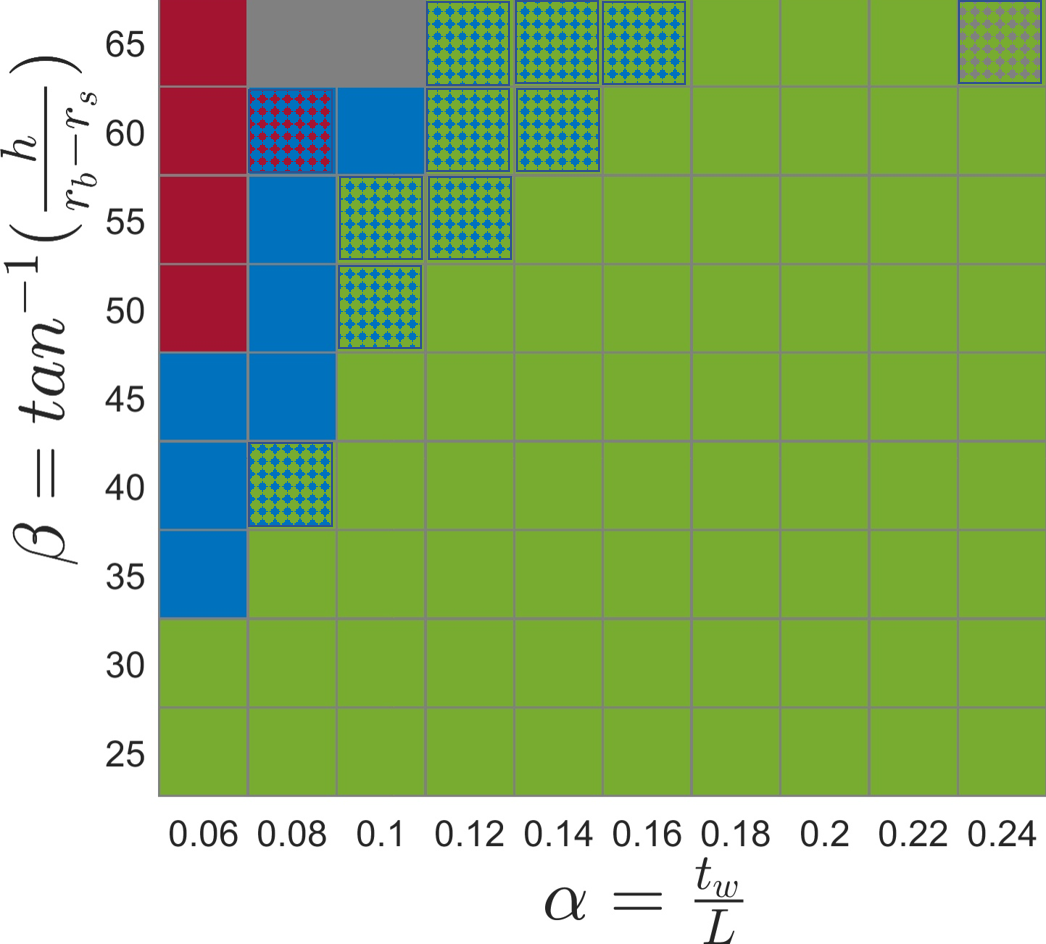} \\
        \midrule
        0.50 & 
        \includegraphics[width=0.4\textwidth]{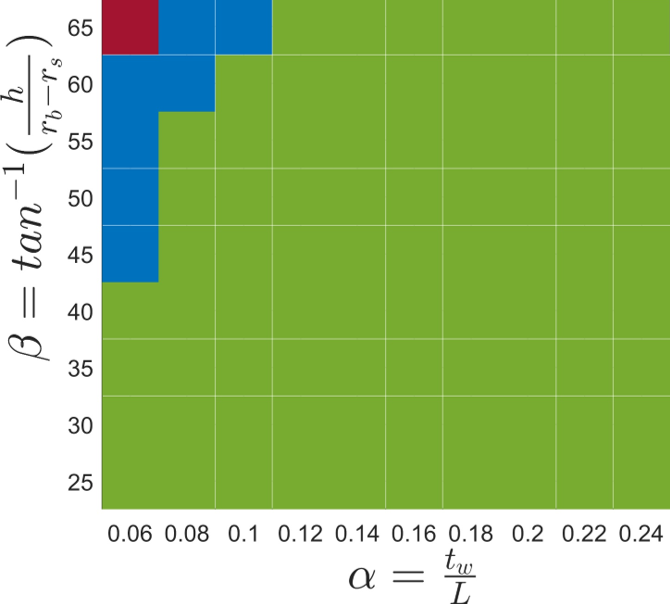} & 
        \includegraphics[width=0.4\textwidth]{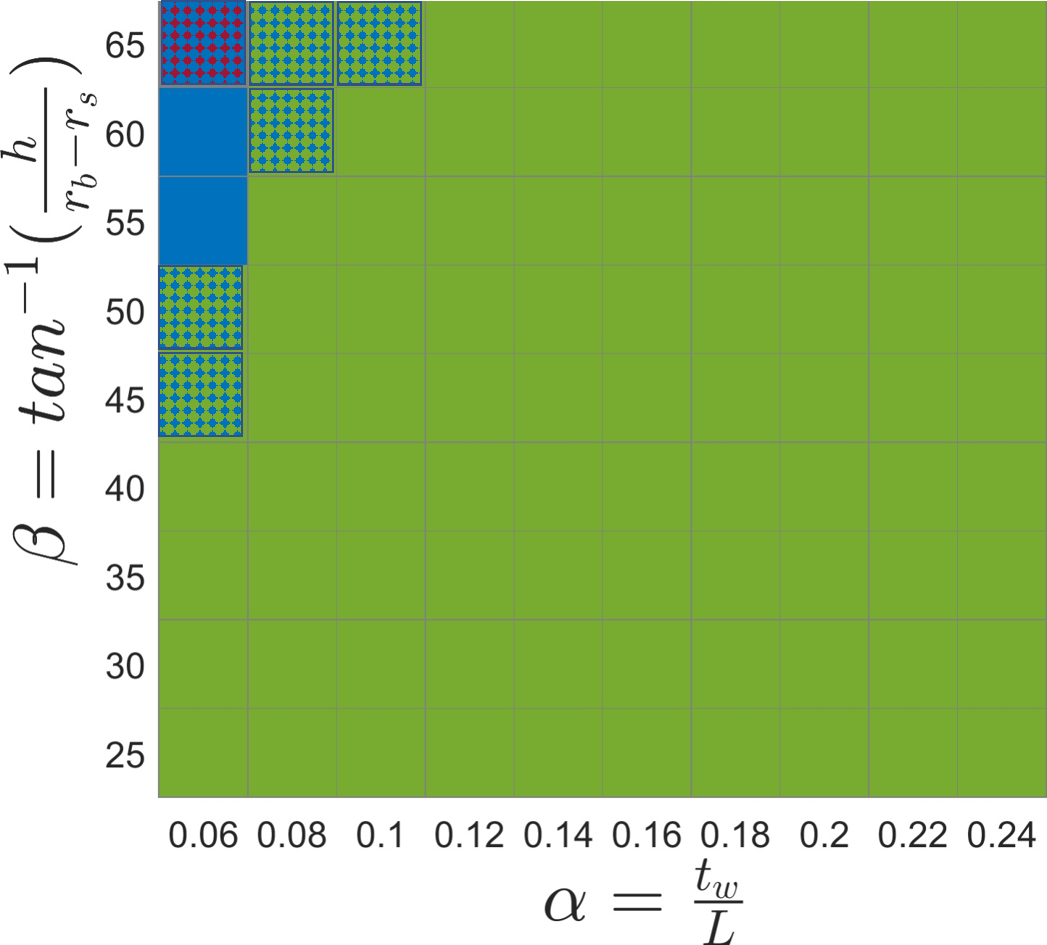} \\
        \midrule
        0.375 & 
        \includegraphics[width=0.4\textwidth]{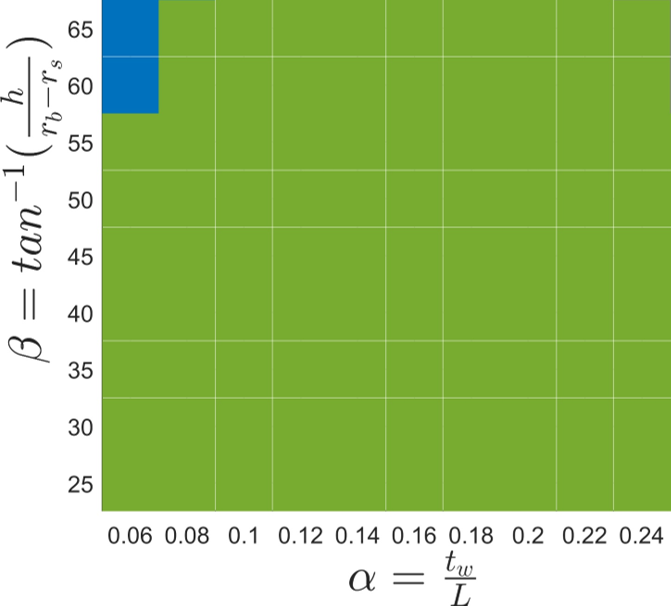} & 
        \includegraphics[width=0.4\textwidth]{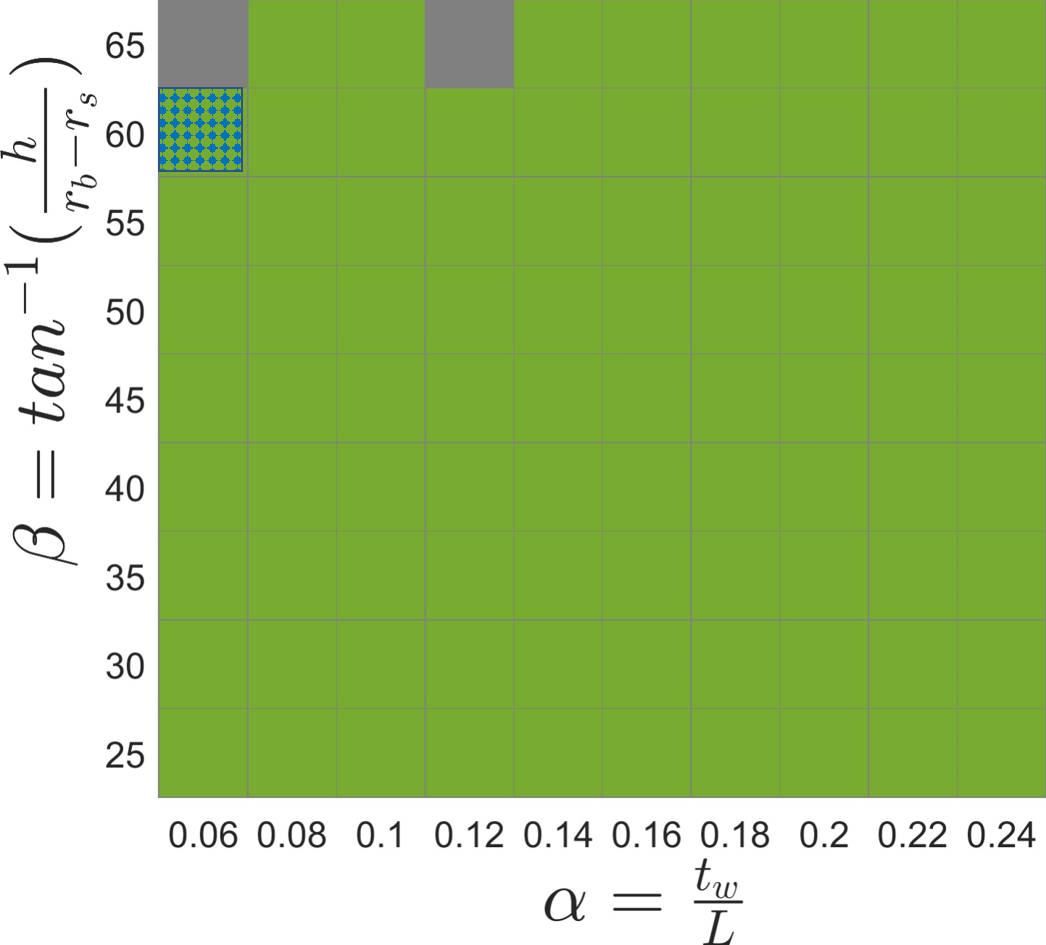} \\
        \midrule
        &\includegraphics[width = 0.65\textwidth]{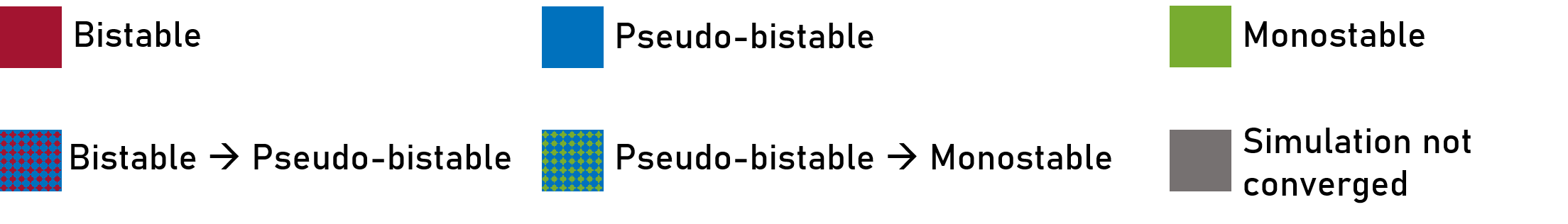} & \\
    \bottomrule
    \end{tabular}
    \caption{Stability conditions for cones with varying \abc\ parametric values. Green represents monostability, red represents bistability, and blue represents pseudo-bistability. The gray regions represent conditions where the simulation did not converge. The right column represents the case when the cones have lateral displacement }
    \label{tab:stability}
\end{table}

\begin{figure}
    \centering
    \includegraphics[width=0.7\textwidth]{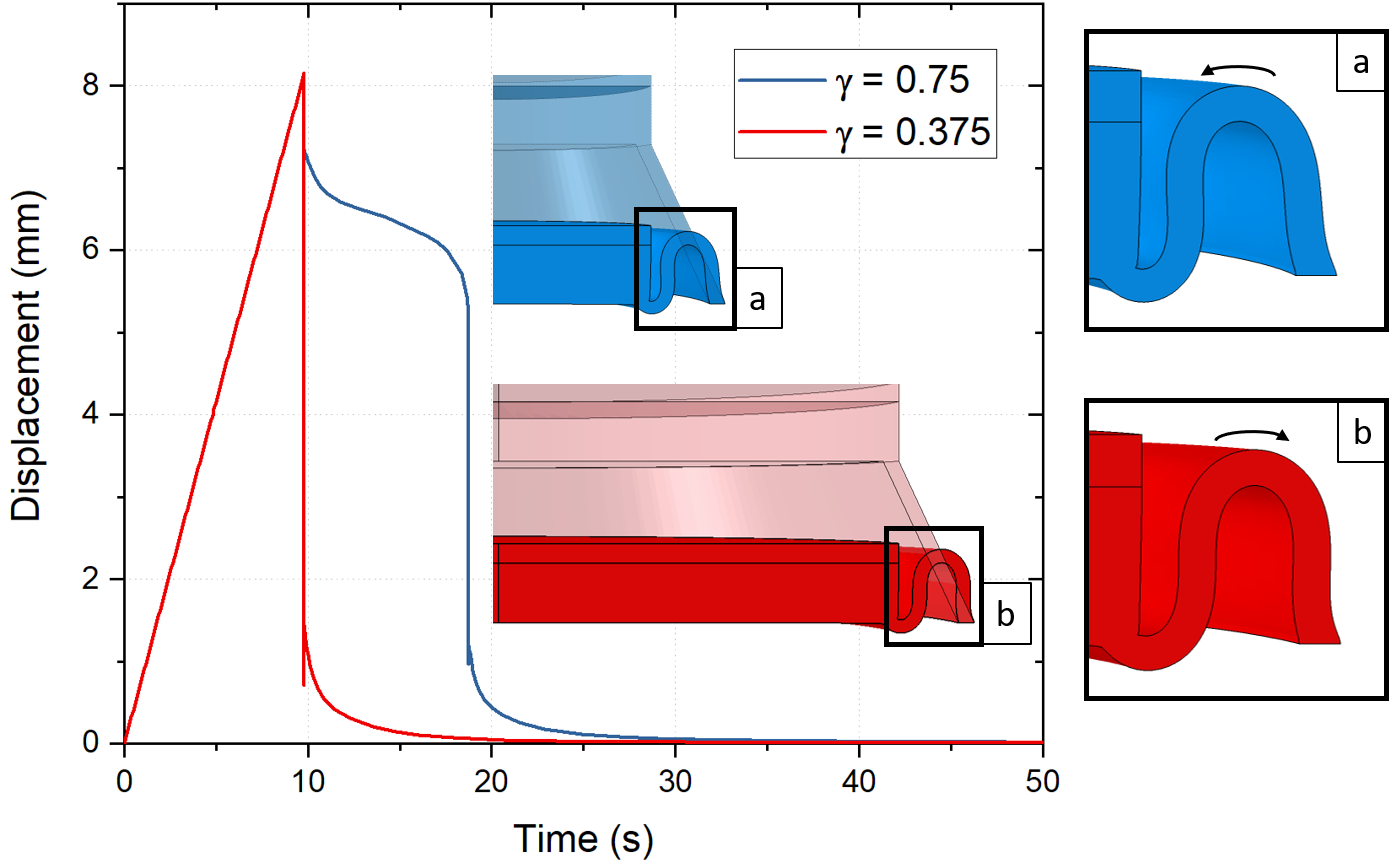}
    \caption{Effect of $\gamma$ on stability: Cones with identical $\alpha, \beta$ = (0.08, 55 \degs) but with different radii of curvatures show different stability conditions. This is due to smaller radius of curvature (larger $\gamma$) resulting in sidewalls of the cone have an inward collapsing effect that causes the cones to exhibit bistability (or pseudo-bistability in this case). Cones with larger radius of curvature have a less pronounced inward collapse that holds the cone in the second configuration. }
    \label{fig:gamma_stability}
\end{figure}

\subsection{Three point bend test results}

\begin{figure}[h]
     \centering
     \begin{subfigure}[b]{0.45\textwidth}
         \centering
         \includegraphics[width=\textwidth]{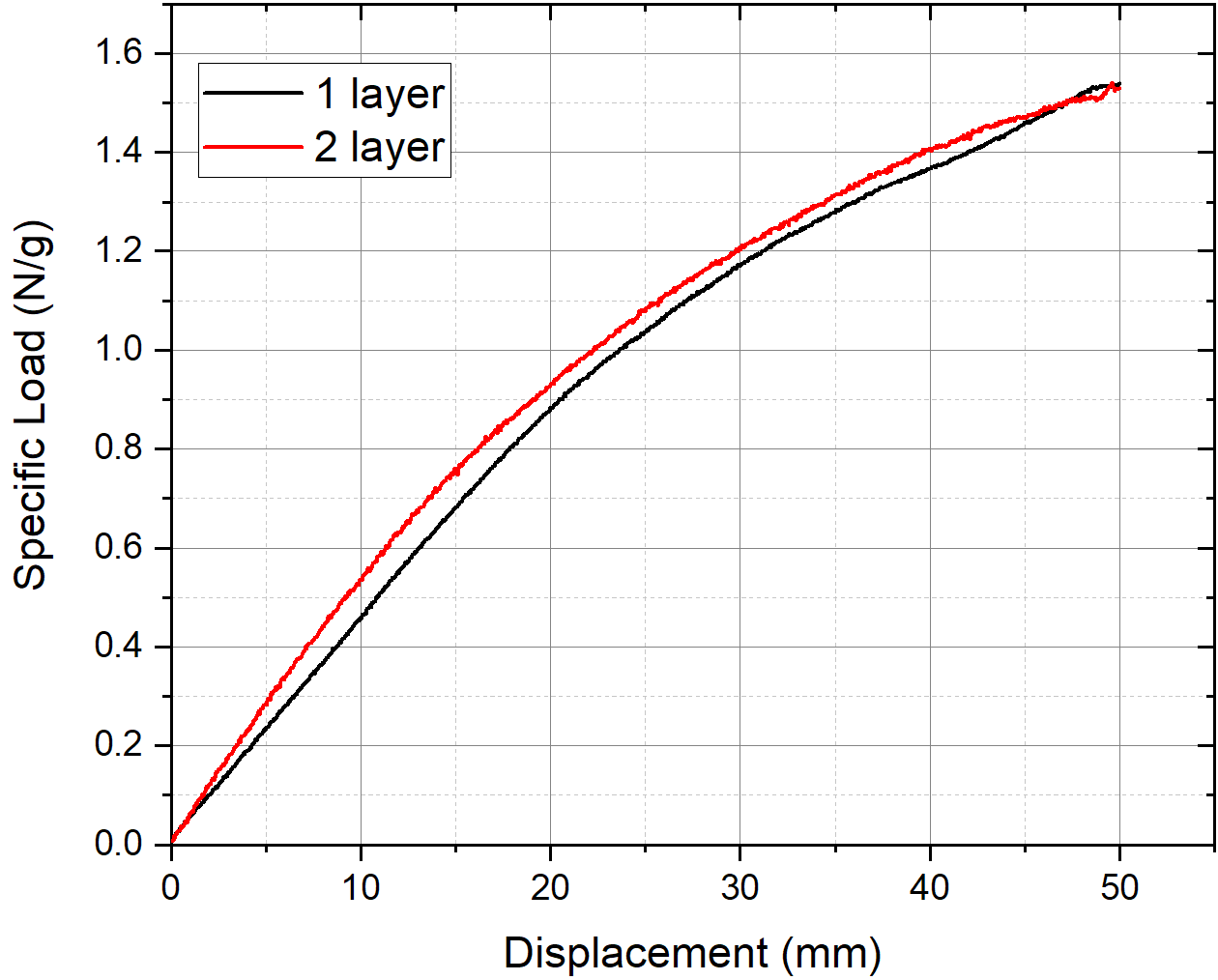}
         \caption{}
         \label{fig:3pt_LD}
     \end{subfigure}
     \hfill
     \begin{subfigure}[b]{0.45\textwidth}
         \centering
         \includegraphics[width=\textwidth]{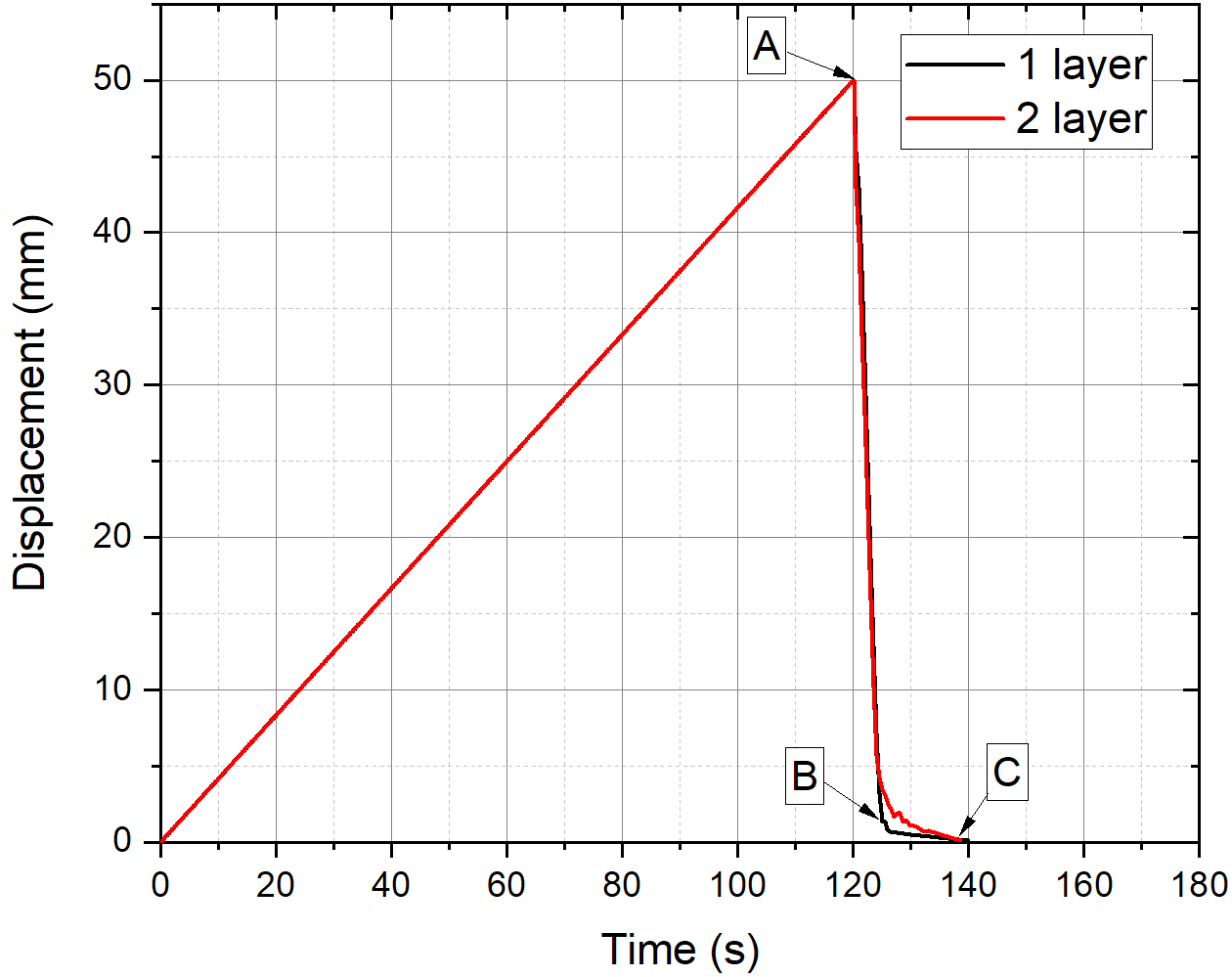}
         \caption{}
         \label{fig:3pt_DispTime}
     \end{subfigure}
     \vfill
     \begin{subfigure}[b]{0.48\textwidth}
         \centering
         \includegraphics[width=\textwidth]{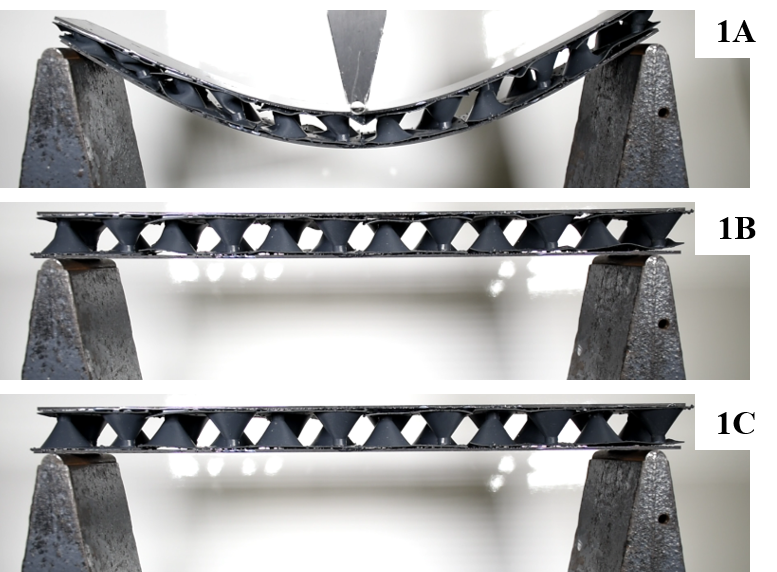}
         \caption{}
         \label{fig:3pt_1L}
     \end{subfigure}
     \hfill
     \begin{subfigure}[b]{0.48\textwidth}
         \centering
         \includegraphics[width=\textwidth]{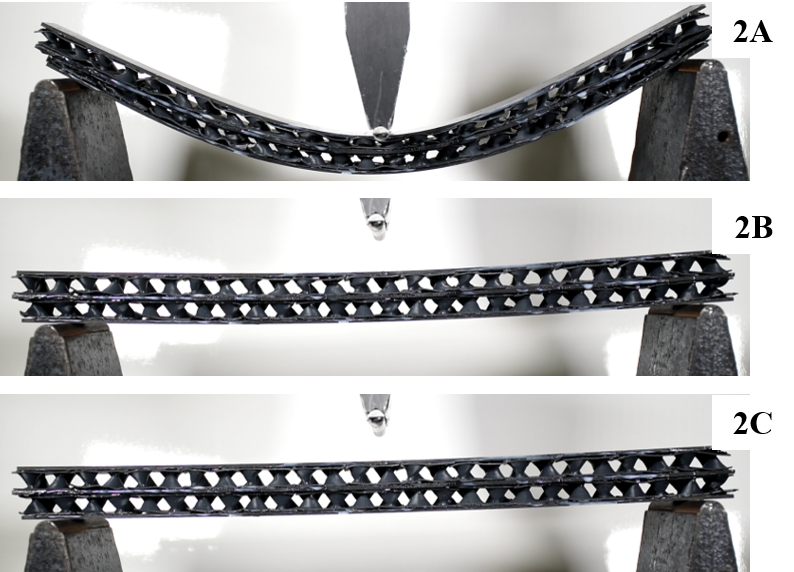}
         \caption{}
         \label{fig:3pt_2L}
     \end{subfigure}
     \caption{Three point bend test results for 1-layer and 2-layer samples with 1mm equivalent wall thickness. a) Load-displacement plot showing the load normalized by weight of the structure. b) Displacement-time plot showing the recovery time of the structure from peak displacement of 50mm. Annotations `A', `B', `C' correspond to stages in the 3-point bending seen in following figures. c) 3-point bending of 1-layer sample, and d) 3-point bending of 2-layer sample. }
     \label{fig:3pt_bend_results}
\end{figure}

Three point bend test results for experiments performed at a loading rate of 25mm/min are shown in Figure~\ref{fig:3pt_bend_results}. The load response was normalized by the weight of the structure as seen in Figure~\ref{fig:3pt_LD}. We see that the weight-normalized load response of the structure in bending remains the same with stacking multiple layers. The displacement-time response recorded by tracking the mid-span point directly under the roller in the video is shown in Figure~\ref{fig:3pt_DispTime}. In this case, all the samples tested showed complete recovery and exhibited monostability. This is attributed to the in-plane shear in the cones away from the mid-span that are pushed towards monostability due to the lateral displacement they experience, as previously shown in Table~\ref{tab:stability}. This is shown in the Figures~\ref{fig:3pt_1L} \&~\ref{fig:3pt_2L}

\section{Conclusions}
\label{se:conclusions}
In this paper, we introduce a method to design and fabricate architected core structures for use as the core in composite sandwich structures. These core structures rely on buckling to dissipate energy during excessive compressive loading. Moreover, due to the use of viscoelastic materials, these core structures exhibit pseudo-bistable behavior that can be exploited to design structures that are not only good energy dissipators, but also show structural recovery without the use of external stimuli.

In summary, \begin{itemize}
    \item The response of truncated cone shaped unit cells were explored as an effect of their geometric properties. Non-dimensional geometric parameters were introduced to characterize the structures, and purely axial ($\epsilon_v = 1, \epsilon_h = 0$), as well as combined axial and shear loading ($\epsilon_v = 1, \epsilon_h = 0.1$) were investigated.
    \begin{itemize}
        \item $\boldsymbol\alpha = \frac{t_w}{L}$ captures the slenderness of the sidewall and determines the buckling load capacity of the structures, where peak buckling load is directly proportional to $\alpha$. 
        \item $\boldsymbol\beta = tan^{-1}(\frac{h}{r_b-r_s})$ captures the incline of the cone's sidewall with respect to its base. A higher $\beta$ makes the cone behavior tend to a cylinder's behavior, making the buckling load higher. 
        \item $\boldsymbol\gamma = \frac{L}{r_b}$ captures the curvature of the sidewall in the $\theta$ or circumferential direction. This $\gamma$ parameter is crucial in determining the post buckling stability of the structures. A higher $\gamma$, or smaller radius of curvature, tends to create a `locking' effect in the structure that forces the structure to exhibit bistability, or pseudo-bistability, depending on the other parameters $\alpha, \beta$.
        \item Introducing a lateral, or shear, displacement in the loading ($\epsilon_h \neq 0$) makes the structures return to their initial undeformed configurations from their deformed configurations. This is attributed to the restoring tensile force on the far side of the cone to the direction of lateral loading vector. This is also shown experimentally in the form of 3-point bending tests, where the cones near the supports experience combined axial-shear loading, and the cones near the center experience axial loading due to symmetry.
    \end{itemize}
    \item The viscoelastic relation of the 3D printed resin introduces an intermediate state of stability between monostable and bistable response, in the form of time-delayed recovery, otherwise known as pseudo-bistability. Due to this pseudo-bistability, structures that are designed to buckle and collapse to dissipate energy without accumulating permanent deformation can recover on their own after a certain period of time, exhibiting structural resilience. 
    \item Increasing the number of layers in the sandwich structure allows us to reduce the size of each unit cell and densely pack the interior region of the core to increase its load carrying capacity or flexural rigidity without accumulating a large weight penalty.
\end{itemize}

\section*{Data Availability}
The raw/processed data required to reproduce these findings cannot be shared at this time as the data also forms part of an ongoing study.

\section*{Acknowledgements}
The authors would like to thank University of Wisconsin-Madison Makerspace for providing 3D printing services and the TeamLab at UW-Madison for waterjet cutting services. We would like to thank Jacob Zeuske from Wisconsin Structures and Materials Testing Lab at UW-Madison for the valuable support for designing the test fixture used in this study, and Kelsey Hacker for conducting preliminary studies to test feasibility of 3D printing flexible architected structures.
We would like to acknowledge the support through the U.S. Office of Naval Research - Young Investigator Program (ONR-YIP) award [N00014-19-1-2206] through {\em{Sea-based Aviation: Structures and Materials Program}} for funding the research presented here.

{\footnotesize
\bibliographystyle{unsrt}
\bibliography{Core-Paper}

\begin{thebibliography}{10}

\bibitem{gupta_characterization_2005}
Nikhil Gupta and Eyassu Woldesenbet.
\newblock Characterization of {Flexural} {Properties} of {Syntactic} {Foam}
  {Core} {Sandwich} {Composites} and {Effect} of {Density} {Variation}.
\newblock {\em Journal of Composite Materials}, 39(24):2197--2212, December
  2005.

\bibitem{breunig_dynamic_2020}
P~Breunig, V~Damodaran, K~Shahapurkar, S~Waddar, M~Doddamani, P~Jeyaraj, and
  P~Prabhakar.
\newblock Dynamic impact behavior of syntactic foam core sandwich composites.
\newblock {\em Journal of Composite Materials}, 54(4):535--547, February 2020.

\bibitem{rizov_indentation_2005}
V.~Rizov, A.~Shipsha, and D.~Zenkert.
\newblock Indentation study of foam core sandwich composite panels.
\newblock {\em Composite Structures}, 69(1):95--102, June 2005.

\bibitem{bai_compression_2019}
Ruixiang Bai, Jingjing Guo, Zhenkun Lei, Da~Liu, Yu~Ma, and Cheng Yan.
\newblock Compression after impact behavior of composite foam-core sandwich
  panels.
\newblock {\em Composite Structures}, 225:111181, October 2019.

\bibitem{pathipaka_damage_2018}
Ranjith~Kumar Pathipaka, Kiran~Kumar Namala, Nagasrisaihari Sunkara, and
  Chennakesava~Rao Bandaru.
\newblock Damage characterization of sandwich composites subjected to impact
  loading:.
\newblock {\em Journal of Sandwich Structures \& Materials}, August 2018.

\bibitem{steeves_collapse_2004}
Craig~A. Steeves and Norman~A. Fleck.
\newblock Collapse mechanisms of sandwich beams with composite faces and a foam
  core, loaded in three-point bending. {Part} {II}: experimental investigation
  and numerical modelling.
\newblock {\em International Journal of Mechanical Sciences}, 46(4):585--608,
  April 2004.

\bibitem{buitrago_modelling_2010}
Brenda~L. Buitrago, Carlos Santiuste, Sonia Sánchez-Sáez, Enrique Barbero,
  and Carlos Navarro.
\newblock Modelling of composite sandwich structures with honeycomb core
  subjected to high-velocity impact.
\newblock {\em Composite Structures}, 92(9):2090--2096, August 2010.

\bibitem{belouettar_experimental_2009}
S.~Belouettar, A.~Abbadi, Z.~Azari, R.~Belouettar, and P.~Freres.
\newblock Experimental investigation of static and fatigue behaviour of
  composites honeycomb materials using four point bending tests.
\newblock {\em Composite Structures}, 87(3):265--273, February 2009.

\bibitem{sun_structural_2021}
Guangyong Sun, Xintao Huo, Hongxu Wang, Paul~J. Hazell, and Qing Li.
\newblock On the structural parameters of honeycomb-core sandwich panels
  against low-velocity impact.
\newblock {\em Composites Part B: Engineering}, 216:108881, July 2021.

\bibitem{meo_response_2005}
M.~Meo, R.~Vignjevic, and G.~Marengo.
\newblock The response of honeycomb sandwich panels under low-velocity impact
  loading.
\newblock {\em International Journal of Mechanical Sciences}, 47(9):1301--1325,
  September 2005.

\bibitem{cricri_honeycomb_2013}
G.~Cricrì, M.~Perrella, and C.~Calì.
\newblock Honeycomb failure processes under in-plane loading.
\newblock {\em Composites Part B: Engineering}, 45(1):1079--1090, February
  2013.

\bibitem{hayes_mechanics_2004}
Alethea~M. Hayes, Aijun Wang, Benjamin~M. Dempsey, and David~L. McDowell.
\newblock Mechanics of linear cellular alloys.
\newblock {\em Mechanics of Materials}, 36(8):691--713, August 2004.

\bibitem{Shan2015}
Sicong Shan, Sung~H Kang, Jordan~R Raney, Pai Wang, Lichen Fang, Francisco
  Candido, Jennifer~A Lewis, and Katia Bertoldi.
\newblock {Multistable Architected Materials for Trapping Elastic Strain
  Energy}.
\newblock {\em Advanced Materials}, 27(29):4296--4301, aug 2015.

\bibitem{tan_novel_2019}
Xiaojun Tan, Bing Wang, Kaili Yao, Shaowei Zhu, Shuai Chen, Peifei Xu, Lianchao
  Wang, and Yuguo Sun.
\newblock Novel multi-stable mechanical metamaterials for trapping energy
  through shear deformation.
\newblock {\em International Journal of Mechanical Sciences}, 164:105168,
  December 2019.

\bibitem{chen_novel_2020}
Shuai Chen, Bing Wang, Shaowei Zhu, Xiaojun Tan, Jiqiang Hu, Xu~Lian, Lianchao
  Wang, and Linzhi Wu.
\newblock A novel composite negative stiffness structure for recoverable
  trapping energy.
\newblock {\em Composites Part A: Applied Science and Manufacturing},
  129:105697, February 2020.

\bibitem{dorfmeister_static_2018}
M.~Dorfmeister, M.~Schneider, and U.~Schmid.
\newblock Static and dynamic performance of bistable {MEMS} membranes.
\newblock {\em Sensors and Actuators A: Physical}, 282:259--268, October 2018.

\bibitem{capanu_design_2000}
M.~Capanu, J.G. Boyd, and P.J. Hesketh.
\newblock Design, fabrication, and testing of a bistable electromagnetically
  actuated microvalve.
\newblock {\em Journal of Microelectromechanical Systems}, 9(2):181--189, June
  2000.

\bibitem{casals-terre_snap-action_2008}
Jasmina Casals-Terré, Andreu Fargas-Marques, and Andrei~M. Shkel.
\newblock Snap-action bistable micromechanisms actuated by nonlinear resonance.
\newblock {\em Journal of Microelectromechanical Systems}, 17(5):1082--1093,
  2008.

\bibitem{rothemund_soft_2018}
Philipp Rothemund, Alar Ainla, Lee Belding, Daniel~J. Preston, Sarah Kurihara,
  Zhigang Suo, and George~M. Whitesides.
\newblock A soft, bistable valve for autonomous control of soft actuators.
\newblock {\em Science Robotics}, 3(16), March 2018.

\bibitem{schomburg_design_1998}
W.~K. Schomburg and C.~Goll.
\newblock Design optimization of bistable microdiaphragm valves.
\newblock {\em Sensors and Actuators A: Physical}, 64(3):259--264, January
  1998.
\newblock Publisher: Elsevier.

\bibitem{goll_microvalves_1996}
C.~Goll, W.~Bacher, B.~Büstgens, D.~Maas, W.~Menz, and W.~K. Schomburg.
\newblock Microvalves with bistable buckled polymer diaphragms.
\newblock {\em Journal of Micromechanics and Microengineering}, 6(1):77--79,
  March 1996.
\newblock Publisher: IOP Publishing.

\bibitem{uusitalo_novel_2010}
Jukka-Pekka Uusitalo, Ville Ahola, Lasse Soederlund, Matti Linjama, Maarit
  Juhola, and Lauri Kettunen.
\newblock Novel {Bistable} {Hammer} {Valve} {For} {Digital} {Hydraulics}.
\newblock {\em International Journal of Fluid Power}, 11(3), January 2010.

\bibitem{pirrera_bistable_2010}
A.~Pirrera, D.~Avitabile, and P.~M. Weaver.
\newblock Bistable plates for morphing structures: {A} refined analytical
  approach with high-order polynomials.
\newblock {\em International Journal of Solids and Structures},
  47(25):3412--3425, December 2010.

\bibitem{wang_bistable_2015}
Bing Wang and Kevin~S. Fancey.
\newblock A bistable morphing composite using viscoelastically generated
  prestress.
\newblock {\em Materials Letters}, 158:108--110, November 2015.

\bibitem{zhang_bistable_2019}
Zheng Zhang, Yang Li, Xiaochen Yu, Xianghao Li, Helong Wu, Huaping Wu, Shaofei
  Jiang, and Guozhong Chai.
\newblock Bistable morphing composite structures: {A} review.
\newblock {\em Thin-Walled Structures}, 142:74--97, September 2019.

\bibitem{chen_nonlinear_2012}
Zi~Chen, Qiaohang Guo, Carmel Majidi, Wenzhe Chen, David~J. Srolovitz, and
  Mikko~P. Haataja.
\newblock Nonlinear {Geometric} {Effects} in {Mechanical} {Bistable} {Morphing}
  {Structures}.
\newblock {\em Physical Review Letters}, 109(11):114302, September 2012.
\newblock Publisher: American Physical Society.

\bibitem{schultz_concept_2008}
Marc~R. Schultz.
\newblock A {Concept} for {Airfoil}-like {Active} {Bistable} {Twisting}
  {Structures}.
\newblock {\em Journal of Intelligent Material Systems and Structures},
  19(2):157--169, February 2008.
\newblock Publisher: SAGE Publications Ltd STM.

\bibitem{forterre_how_2005}
Yoël Forterre, Jan~M. Skotheim, Jacques Dumais, and L.~Mahadevan.
\newblock How the {Venus} flytrap snaps.
\newblock {\em Nature}, 433(7024):421--425, January 2005.

\bibitem{alturki_multistable_2019}
Mansour Alturki and Rigoberto Burgueño.
\newblock Multistable cosine-curved dome system for elastic energy dissipation.
\newblock {\em Journal of Applied Mechanics, Transactions ASME}, 86(9):1--10,
  2019.

\bibitem{cui_highly_2015}
Yuefeng Cui and Matthew Santer.
\newblock Highly multistable composite surfaces.
\newblock {\em Composite Structures}, 124:44--54, June 2015.

\bibitem{restrepo_phase_2015}
David Restrepo, Nilesh~D. Mankame, and Pablo~D. Zavattieri.
\newblock Phase transforming cellular materials.
\newblock {\em Extreme Mechanics Letters}, 4:52--60, 2015.
\newblock ISBN: 2352-4316 Publisher: Elsevier Ltd.

\bibitem{tan_novel_2020}
Xiaojun Tan, Bing Wang, Shaowei Zhu, Shuai Chen, Kaili Yao, Peifei Xu, Linzhi
  Wu, and Yuguo Sun.
\newblock Novel multidirectional negative stiffness mechanical metamaterials.
\newblock {\em Smart Materials and Structures}, 29(1), 2020.

\bibitem{che_viscoelastic_2018}
Kaikai Che, Chao Yuan, H.~Jerry Qi, and Julien Meaud.
\newblock Viscoelastic multistable architected materials with
  temperature-dependent snapping sequence.
\newblock {\em Soft Matter}, 14(13):2492--2499, 2018.
\newblock Publisher: Royal Society of Chemistry.

\bibitem{tao_4d_2020}
Ran Tao, Li~Xi, Wenwang Wu, Ying Li, Binbin Liao, Liwu Liu, Jinsong Leng, and
  Daining Fang.
\newblock {4D} printed multi-stable metamaterials with mechanically tunable
  performance.
\newblock {\em Composite Structures}, 252(May), 2020.

\bibitem{liu_4d_2020}
Kai Liu, Le~Han, Wenxia Hu, Longtao Ji, Shengxin Zhu, Zhishuai Wan, Xudong
  Yang, Yuling Wei, Zongjie Dai, Zeang Zhao, Zhen Li, Pengfei Wang, and Ran
  Tao.
\newblock {4D} printed zero {Poisson}'s ratio metamaterial with switching
  function of mechanical and vibration isolation performance.
\newblock {\em Materials and Design}, 196:109153, 2020.

\bibitem{medina_bistable_2016}
Lior Medina, Rivka Gilat, and Slava Krylov.
\newblock Bistable behavior of electrostatically actuated initially curved
  micro plate.
\newblock {\em Sensors and Actuators A: Physical}, 248:193--198, September
  2016.

\bibitem{seffen_eversion_2016}
Keith~A Seffen and Stefano Vidoli.
\newblock Eversion of bistable shells under magnetic actuation: a model of
  nonlinear shapes.
\newblock {\em Smart Materials and Structures}, 25(6):065010, June 2016.

\bibitem{santer_self-actuated_2010}
M.~Santer.
\newblock Self-actuated snap back of viscoelastic pulsing structures.
\newblock {\em International Journal of Solids and Structures},
  47(24):3263--3271, December 2010.

\bibitem{brinkmeyer_pseudo-bistable_2012}
A.~Brinkmeyer, M.~Santer, A.~Pirrera, and P.~M. Weaver.
\newblock Pseudo-bistable self-actuated domes for morphing applications.
\newblock {\em International Journal of Solids and Structures},
  49(9):1077--1087, 2012.

\bibitem{shan_multistable_2015}
Sicong Shan, Sung~H Kang, Jordan~R Raney, Pai Wang, Lichen Fang, Francisco
  Candido, Jennifer~A Lewis, and Katia Bertoldi.
\newblock Multistable {Architected} {Materials} for {Trapping} {Elastic}
  {Strain} {Energy}.
\newblock {\em Advanced Materials}, 27(29):4296--4301, August 2015.

\bibitem{astm_d11_committee_test_nodate}
ASTM~D11 Committee.
\newblock Test {Methods} for {Vulcanized} {Rubber} and {Thermoplastic}
  {ElastomersTension}.
\newblock Technical report, ASTM International.

\bibitem{MitsubishiRayonCarbonFiber&Composites2014}
{Mitsubishi Rayon Carbon Fiber {\&} Composites}.
\newblock {\em {NB301 Data Sheet}}.
\newblock www.rockwestcomposites.com/media/wysiwyg/ MRCFC{\_}NB301{\_}Data.pdf,
  2014.

\end{thebibliography}
}

\end{document}